\documentclass[aps,prb,showpacs,twocolumn,groupedaddress]{revtex4-1}

\usepackage{amsmath}
\usepackage{amssymb}
\usepackage{graphicx}
\usepackage[english]{babel}
\usepackage{blindtext}
\usepackage[utf8]{inputenc}
\usepackage{color}

\newcommand{\AlTwoOThree}{Al$_{\mathrm{2}}$O$_{\mathrm{3}}$}

\newcommand{\SiOTwo}{SiO$_{\mathrm{2}}$}

\begin{document} 
\title{Electron surface scattering kernel for a plasma facing a semiconductor}

\author{F. X. Bronold and F. Willert} 
\affiliation{Institut f{\"ur} Physik,
             Universit{\"a}t Greifswald,
             17489 Greifswald,
             Germany}

\date{\today}
\begin{abstract}
Employing the invariant embedding principle for the electron backscattering 
function, we present a scheme for constructing an electron surface 
scattering kernel to be used in the boundary condition for the electron Boltzmann 
equation of a plasma facing a semiconducting solid. The scheme 
takes the solid's microphysics responsible for electron emission and backscattering 
from the interface within a randium-jellium model into account and is applicable 
to dielectrics and metals as well. As an illustration, we consider silicon and 
germanium, describing the interface potential by a Schottky barrier and 
including impact ionization across the energy gap as well as scattering on 
phonons and ion cores. The emission yields deduced from the kernel agree 
well enough with measured data to support its use in the electron boundary condition
of a plasma facing silicon or germanium.
\end{abstract}
\pacs{68.49.Jk, 79.20.Hx, 52.40.Hf}
\maketitle

\section{Introduction} 

Man-made plasmas are bounded by condensed matter. Essentially all commercially
exploited technological plasmas~\cite{WKH19} interact with either 
a liquid or a solid. For instance, plasmas for medical applications naturally 
have contact with human cells and hence with a liquid environment, whereas 
plasmas used for surface modification or surface catalysis are in contact
with solids. Solids and plasmas are especially strongly coupled in 
semiconductor-based microdischarges~\cite{MFS18,EPC13,DOL10}, where the 
surface to volume ratio is particularly large, making the interaction with
the solid an integral part of the physical system. Even magnetically confined 
fusion plasmas interact with condensed matter in the divertor region via 
sheaths~\cite{TKR23,TKR20,Campanell20} and dust particles~\cite{VDT18,Tolias14b}, as 
do plasmas employed for electric propulsion in Hall thrusters~\cite{RSK05,DRF03}.   

Although not complete, the list indicates that a kinetic description of 
technological plasmas, based on equations for the electromagnetic fields 
and a set of Boltzmann equations for the various particle species, requires 
boundary conditions for the fields and the particle distribution functions. As 
in any Boltzmann-type modeling of kinetic phenomena~\cite{Williams71,Kogan69,Alpert65}, 
the latter are integrals relating, for each species, at the boundary 
the distribution function of the outgoing particles with that of the incoming ones. 
Hence they control the flux balance at the boundary and have a strong effect 
on the plasma sheath as well as the overall characteristics of the plasma.

The kernel of such an integral--the surface scattering kernel~\cite{Cercignani95}--is a 
complicated object, because it arises from the microscopic processes at or within 
the bounding medium responsible for particle reflection and/or emission. Mathematical 
constraints enforced by the processes, such as positivity, normalization, and--in 
thermal equilibrium--reciprocity~\cite{Kuscer71}, can be straightforwardly formulated, 
but setting up for each species an expression for the kernel, from which 
quantitative data can be deduced, requires to solve the kinetic problem also 
partly within the bounding 
medium. To avoid this task, phenomenological kernels~\cite{CL71}, containing 
a set of adjustable parameters, are widely used. For instance, a two-term scattering 
kernel, describing specular and diffuse reflection, has been employed in neutron 
transport~\cite{Williams71}, gas kinetics~\cite{Kogan69}, as well as plasma 
physics~\cite{Alpert65}. The electron boundary condition most popular in the modeling 
of technological low-temperature plasmas~\cite{BGL10,LSW02,AGF97} even considers only 
specular reflection. It contains the electron reflection probability as an adjustable 
parameter. 

Recently, however, an effort started to determine for plasma-exposed surfaces the 
electron reflection probability and the closely related secondary electron emission 
yield experimentally~\cite{SDB22,DBS16,DAK15}. The material dependence of the two 
parameters moves also more and more in the focus of a quantitative plasma 
modeling~\cite{BS24,CSG23,PHD23,HDS22,DDM19,HDK17,TLC04}. It is thus appropriate to 
set up for electrons boundary conditions containing the wall's microphysics more 
faithfully than the parameterized boundary conditions used so far. An 
example where this matters is the work of Horv\'{a}th and 
coworkers~\cite{HDS22,HDK17}. By particle-in-cell/Monte Carlo collisions simulations, 
they identified regimes in capacitively coupled radio frequency discharges where the 
ionization dynamics is strongly affected by the angle and energy dependence of the 
electron-induced secondary electron emission coefficient. Knowing this quantity and 
including it realistically into simulations is thus critical for a deeper 
understanding of this type of discharge, which is used in many technological 
plasma applications.

The purpose of this work is to construct a physical boundary condition for the electron
Boltzmann equation of a plasma in contact with a solid. To illustrate the approach, 
we consider a planar semiconducting plasma-solid interface, as it occurs in 
semiconductor-based microdischarges~\cite{MFS18,EPC13,DOL10}. Instead of resolving  
the electron kinetics inside the solid by a separate Boltzmann equation, we employ the 
invariant embedding principle~\cite{Dashen64,GP07,ARK20} to set up an integral 
equation for the backscattering function which is closely related to the surface 
scattering kernel. We employed this approach before to calculate, at low energies, the 
electron sticking coefficient for dielectrics~\cite{BF15} and the secondary electron 
emission yield~\cite{BF22} for metals. Using the backscattering function derived 
by one of the authors in a previous work~\cite{BF15}, Cagas and coworkers~\cite{CHS20}
also set up a boundary condition for the electron Boltzmann equation in the manner 
we propose in this work. Their implementation did however not include the internal 
scattering cascades. Moreover, they mainly discussed numerical issues of the boundary 
condition, whereas we concentrate on its physics.

Based on the invariant embedding principle~\cite{Dashen64,GP07,ARK20} and a numerical
strategy for its handling developed in nuclear reactor theory~\cite{SM66a,SA72}, we 
compute below an electron surface scattering kernel for a semiconducting interface. 
In its course we also remedy shortages of our previous work~\cite{BF15,BF17a,BFP18,BF22}. 
The kernel includes impact ionization across the band gap~\cite{Kane67,TPE91,BHI92,CFE93} 
and is thus also valid for impact energies larger than the band gap. The electron multiplication 
associated with impact ionization required a renewed analysis of the normalization of the 
backscattering function. Thereby we realized that the normalization used so 
far~\cite{BF15,BF17a,BFP18,BF22} cannot be correct, despite the reasonable sticking and 
emission coefficients it led to, because it gives in the limit of vanishing interface 
potential and particle-number conserving scattering processes an energy and angle 
independent emission yield of exactly one. Moreover, the work on metals~\cite{BF22} suggests, 
that scattering on the ion cores, which we initially thought not to be of importance at the 
low electron impact energies typical for plasma applications, has to be  
also included for dielectrics and semiconductors. Assuming, as for metals, the scattering 
on the cores to be incoherent, we employ for that purpose a randium-jellium 
model~\cite{Bauer70,DMB82}, distributing screened~\cite{Srinivasan69,Phillips68,Penn62}
pseudopotentials~\cite{ILC78,SCL75,Phillips73} randomly within the solid. Finally, the 
interface potential contains now also the image-charge effect~\cite{MacColl39}, which 
significantly reduces the emission yield at very low impact energies. Since the emission 
yields we obtain for silicon and germanium are in reasonable agreement with 
measured~\cite{FF58,BronFrai69} as well as Monte Carlo simulation data~\cite{PIB17}, 
we consider the model employed in this work more reliable than the one used 
before~\cite{BF15}. Using it in the surface scattering kernel should thus lead to 
plausible electron boundary conditions for plasmas in contact with semiconductors or 
dielectrics. 

The remainder of the paper is organized as follows. In Sect.~\ref{Formalism},
divided into three subsections, we present the formalism used for constructing 
a boundary condition for the electron Boltzmann equation of the plasma. 
First, in subsection~\ref{SurfaceScatteringKernel}, we define the surface 
scattering kernel in terms of the backscattering and transmission functions of 
the plasma-solid interface and relate the kernel to the incoming and outgoing
electron energy distribution functions. Subsection~\ref{BackScatteringFunction}
presents our approach for computing the backscattering function from 
the embedding equation without approximation except the discretization of the 
integrals over energy and direction cosine. Finally, in subsection~\ref{Model}, 
the randium-jellium model is introduced. Numerical results for the surface scattering 
kernel and the emission yield are presented in Sect.~\ref{Results} before we 
conclude in Sect.~\ref{Conclusions}. Details interrupting the 
flow of presentation are presented in three appendices. 

\section{Formalism}
\label{Formalism}

Instead of characterizing electron-surface interaction at the plasma-solid interface
by empirical formulae for electron reflection, backscattering, and emission
probabilities/coefficients~\cite{HDS22,HDK17,CSG23,TLC04,BS24}, originating either from the
Vaughan~\cite{Vaughan89} or the Furman-Pivi~\cite{FP02} parameterizations of
secondary electron emission, our approach characterizes the interaction by physical
processes taking place at or inside the solid. It thus has a clear and transparent
physical basis. Moreover, it
provides without further ado the angle- and energy-resolved spectrum of the
electrons coming back to the plasma as a result of electron bombardment of the
surface. In order to obtain an approach flexible enough to be--in its general
structure--applicable to different materials, it is based on a generic model
of the solid. In the present work, we restrict the construction of the electron
surface scattering kernel to semiconducting solids and impact energies below a few
10s eV, where the excitation of plasmons and core electrons can be ignored. However,
the formalism can be applied to other materials and higher energies as well. Different
sets of scattering processes have then to be considered.

\subsection{Surface scattering kernel}
\label{SurfaceScatteringKernel}
\begin{figure*}[t]
\begin{minipage}{0.33\linewidth}
\includegraphics[width=0.90\linewidth]{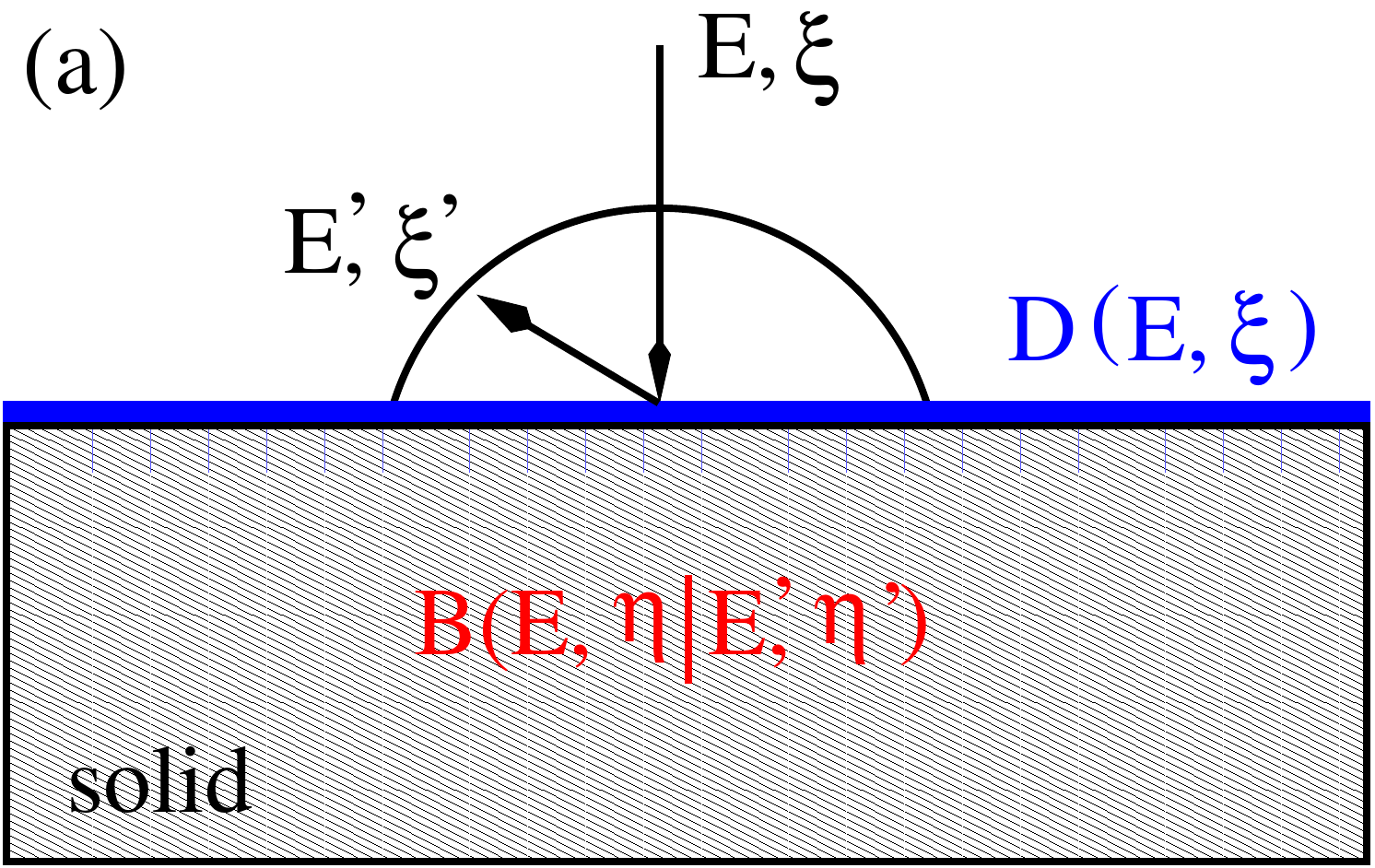}
\end{minipage}\begin{minipage}{0.33\linewidth}
\includegraphics[width=0.90\linewidth]{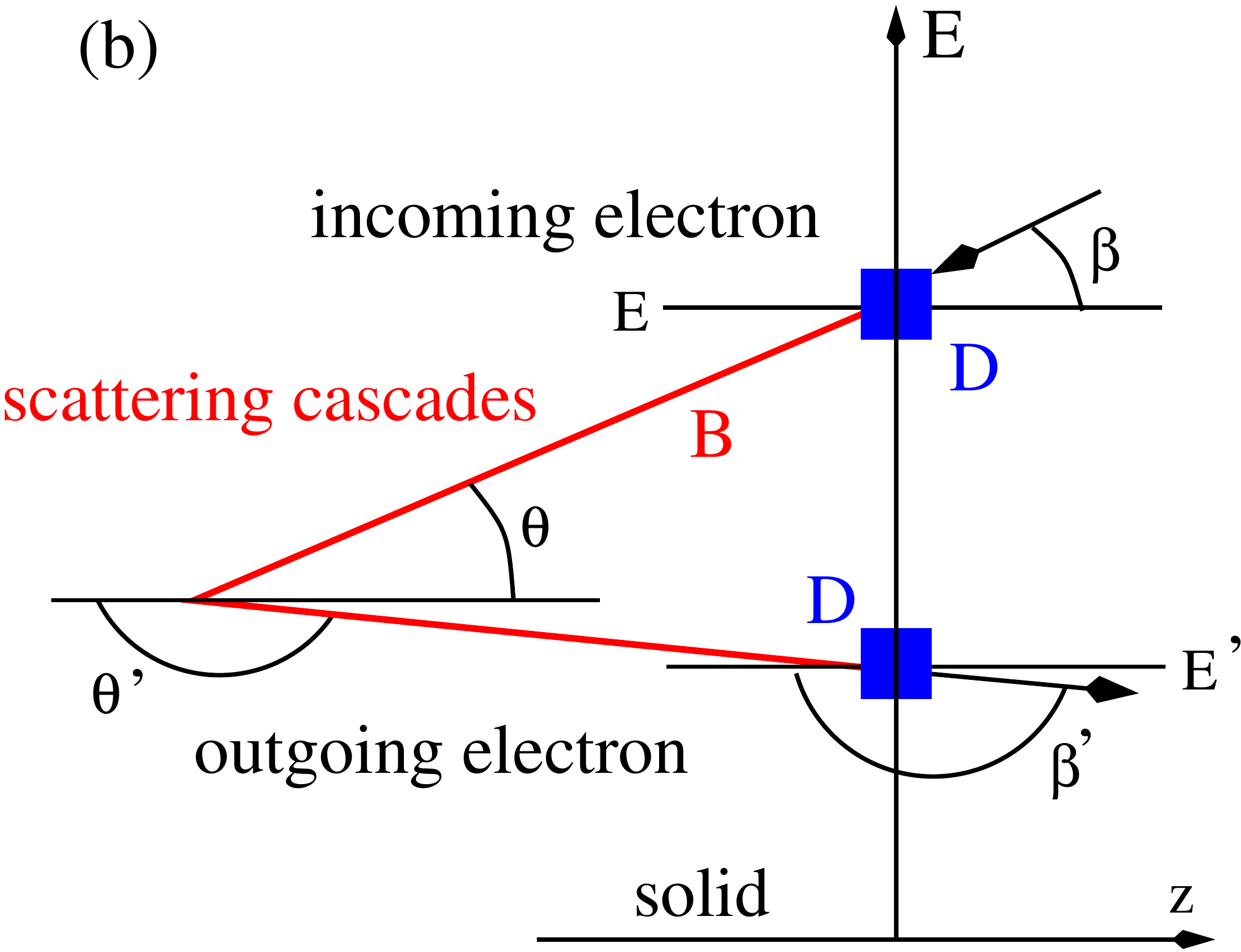}
\end{minipage}\begin{minipage}{0.33\linewidth}
\hfill\includegraphics[width=0.90\linewidth]{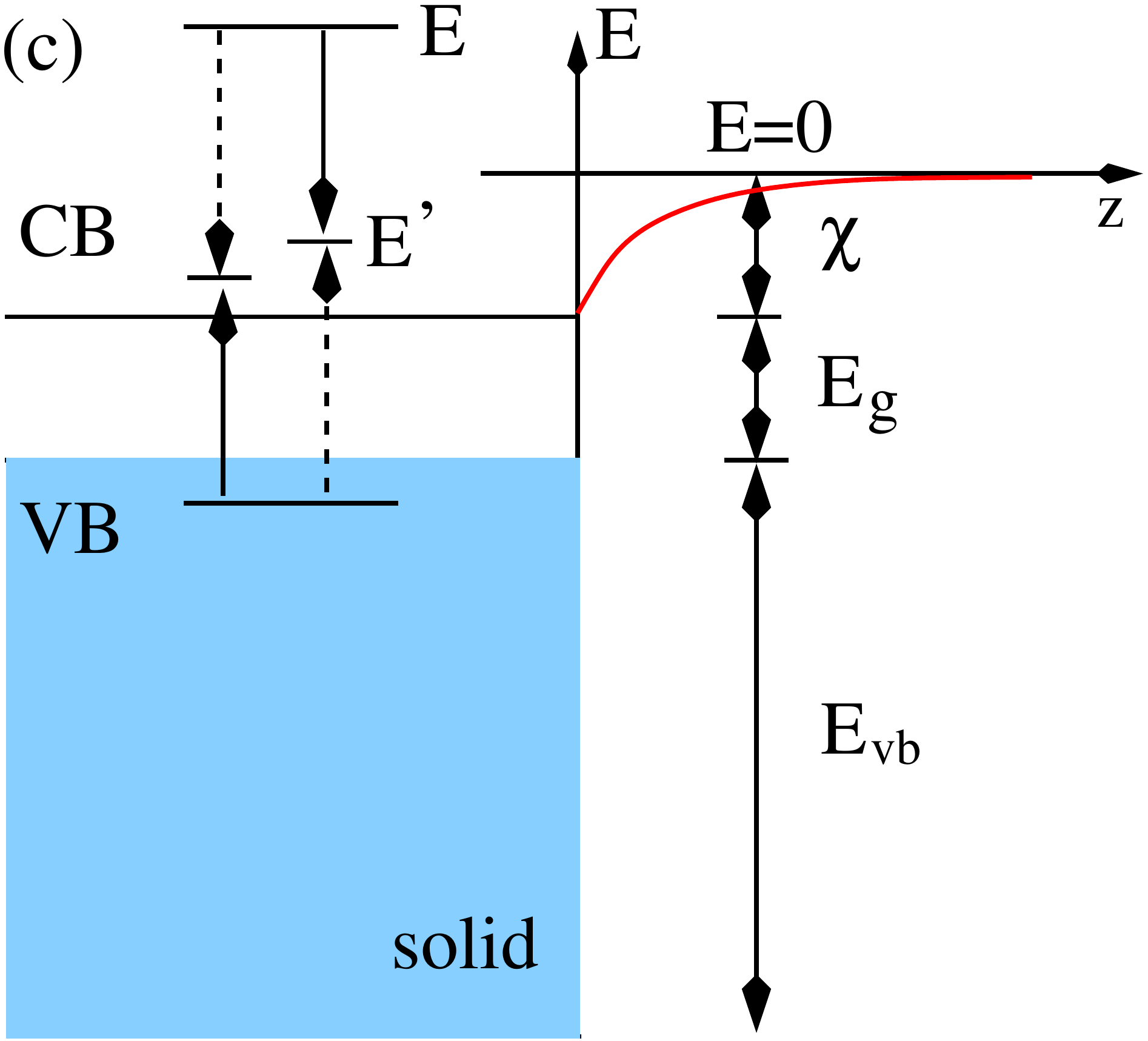}
\end{minipage}
\caption{(Color online) (a) A primary electron with energy $E$ (measured with respect to 
the just-outside potential energy) and direction cosine $\xi$ hits 
a solid and leads to a secondary electron with energy $E^\prime$ and direction cosine $\xi^\prime$.
Both have to traverse the interface potential, modelled by a Schottky barrier, giving rise to the
surface transmission function $D(E,\xi)$. The scattering cascades inside the
solid are encoded in the backscattering function $B(E,\eta|E^\prime,\eta^\prime)$. 
(b) Scattering angles $\beta, \beta^\prime$ and $\theta, \theta^\prime$ used in the definition
of the direction cosines $\xi, \xi^\prime$ and $\eta, \eta^\prime$ outside and inside the solid,
respectively. (c) Two-band model used for the description of the semiconductor. The energies
defining it are the electron affinity $\chi$, the band gap $E_g$, and the width
of the valence band $E_{\rm vb}$ as obtained from the density of the valence electrons.
Also indicated is the Schottky barrier (in red) and an energy loss process due to direct 
and exchange impact ionization.}
\label{ModelCartoon}
\end{figure*}

The quintessence of our approach is summarized in Fig.~\ref{ModelCartoon}a. It 
shows an electron with energy $E$ and direction cosine $\xi$ impinging onto a
laterally homogeneous planar interface at $z=0$ and initiating an outgoing 
electron with energy $E^\prime$ and direction cosine $\xi^\prime$. The motion 
of the electrons is described by their total energy, measured with respect to 
the potential just outside the interface (just-outside potential), their direction 
cosines with respect to the outgoing normal of the interface, which also defines 
the $z-$axis of the coordinate system in real space, and an azimuth angle $\Phi$, 
which due to the lateral homogeneity of the interface can be however integrated out. 
Hence the function $D(E,\xi)$, describing the transmission through the interface 
potential, and the function $B(E,\eta|E^\prime,\eta^\prime)$, encoding the scattering 
cascades inside the solid leading to outgoing electrons, are only functions of 
energy and direction cosine. 

Measuring energy, length, and mass, respectively, in Rydberg energies, Bohr radii, and bare
electron masses, and using the notation introduced in Fig.~\ref{ModelCartoon}b, the surface
scattering kernel $R(E,\xi|E^\prime,\xi^\prime)$ relating at $z=0$ the electron energy 
distribution function $F^<(E,\xi)$ of the incoming electrons with the distribution function 
$F^>(E^\prime,\xi^\prime)$ of the outgoing ones is defined by~\cite{CHS20} 
\begin{align}
F^>(E^\prime,\xi^\prime)=\int_0^\infty dE \int_0^1 d\xi F^<(E,\xi) R(E,\xi|E^\prime,\xi^\prime)
\label{IntegralRelation}
\end{align}
with $E^\prime \geq 0$ and 
\begin{align}
R(E,\xi|E^\prime,\xi^\prime)&=R(E,\xi)\delta(E-E^\prime)\delta(\xi-\xi^\prime)\nonumber\\
&+\Delta R(E,\xi|E^\prime,\xi^\prime)~,
\label{SSK}
\end{align}
where $R(E,\xi)=1-D(E,\xi)$ is the probability for an electron with energy $E$ and 
direction cosine $\xi$ to be quantum-mechanically reflected by the interface potential 
and $\Delta R(E,\xi|E^\prime,\xi^\prime)$ is the part of the kernel accounting for the 
scattering cascades inside the solid producing a backscattered electron with energy 
$0<E^\prime \leq E$ and direction cosine $\xi^\prime$. 

The integral relation~\eqref{IntegralRelation} holds for that part of the incoming electron 
energy distribution function which describes electrons with energy large enough to overcome 
the repulsive wall potential. Only this group of electrons hits the material interface and 
is not reflected by the wall potential. Since we measure energy from the potential just 
outside the interface, electrons of this group have positive energy. Likewise, 
ignoring tunneling through the total potential barrier arising from the matching of the Schottky 
barrier with the repulsive wall potential, which is of minor importance for not too
strong electric fields at the plasma-solid interface, electrons leaving the solid require a 
kinetic energy perpendicular to the interface larger than the electron affinity $\chi$. 
Their total energy $E^\prime$ is thus also positive. Hence, the surface scattering 
kernel~\eqref{SSK} is defined for $E, E^\prime \geq 0$. Situations where tunneling 
matters, for instance, in the multi-emissive sheaths of the divertor regions of fusion 
devices~\cite{TKR23,TKR20,Campanell20} are outside the scope of the present work. Our 
focus is instead on low-temperature plasmas facing a solid without affecting the emissive 
properties of the interface itself.

Since the scattering cascade encoded in $B(E,\eta|E^\prime,\eta^\prime)$ involves states far 
away from the extremal points of the conduction band, we do not employ effective electron 
masses inside the solid as we did before~\cite{BF15,BF17a,BFP18}. Instead, we now simply
take bare electron masses (also for the holes). This is of course an approximation, but 
overcoming it requires to account for the full band structure, including nonparabolicities 
as well as multi-valleys, which is beyond our present scope. It is also not obvious, to 
what extent band structure details affect the surface scattering kernel and the emission 
yield quantitatively. It is conceivable that many band structure details are washed out 
due to the cascade's multiple scattering. 
Hence, the conduction band density of states reads $\rho(E)=\sqrt{E+\chi}/2$
and the relation between the internal ($\eta$) and external direction cosines ($\xi$), 
to be obtained from the conservation of the lateral momentum and the total energy, 
becomes $1-\eta^2=(1-\xi^2)E/(E+\chi)$ from which $\eta(\xi)$ and its inverse $\xi(\eta)$ 
follow. Using, 
$\partial \eta^\prime/\partial\xi^\prime = E^\prime \xi^\prime /((E^\prime + \chi)\eta^\prime)$ 
we finally obtain~\cite{BFP18} 
\begin{align}
\Delta R(E,\xi|E^\prime,\xi^\prime)&=\frac{E^\prime}{E^\prime+\chi}\frac{\xi^\prime}{\eta^\prime}
\rho(E^\prime)\Theta(E-E^\prime)D(E,\xi)\nonumber\\
&\times B(E,\eta(\xi)|E^\prime,\eta(\xi^\prime))D(E^\prime,\xi^\prime)~,
\label{DeltaR}
\end{align}
where the Heaviside step function $\Theta(E-E^\prime)$ ensures $E \geq E^\prime$ and 
$B(E,\eta|E^\prime,\eta^\prime)$ is the backscattering function to which we turn in the 
next subsection. Before, however, we note that in terms of the surface scattering kernel, 
the emission yield is given by
\begin{align}
Y(E,\xi)=\int_0^E dE^\prime \int_0^1 d\xi^\prime R(E,\xi|E^\prime,\xi^\prime)~.
\label{EscapeFct}
\end{align}
It can be transformed into the expression for $Y(E,\xi)$ given before~\cite{BF15,BF17a,BFP18,BF22}
by changing the integration variables from $(E^\prime,\xi^\prime)$ to $(E^\prime,\eta^\prime)$
and taking into account that only internal backscattered states with perpendicular kinetic 
energy larger than the electron affinity contribute to the emission yield.\\

\subsection{Backscattering function}
\label{BackScatteringFunction}

The central object of our approach is the backscattering function $B(E,\eta|E^\prime,\eta^\prime)$. 
It describes the (pseudo-)probability for an electron with energy $E$ and direction cosine $\eta$ 
to lead to a backscattered electron with energy $E^\prime$ and direction cosine $\eta^\prime$. In 
the previous work~\cite{BF15,BF17a,BFP18,BF22}, we considered $B(E,\eta|E^\prime,\eta^\prime)$ 
as a conditional probability, obtained from the function 
$Q(E,\eta|E^\prime,\eta^\prime)$ normalized to the totality of all backscattered states, 
including those, which do not lead to electron escape from the solid. The emission yields we 
obtained turned out to be in good agreement with experimental data suggesting that 
the normalization is indeed required. However, in the course of the present investigation 
we realized that the normalization leads in the limit of particle-conserving scattering 
processes and vanishing work function (metal) or electron affinity (semiconductor) to an energy 
and angle independent unitary emission yield. Although in reality not realizable, this cannot be 
correct. In the following, we therefore abandon the conditional probability construction and 
identify the backscattering function $B(E,\eta|E^\prime,\eta^\prime)$ directly with the function 
$Q(E,\eta|E^\prime,\eta^\prime)$, that is, we set
\begin{align}
B(E,\eta|E^\prime,\eta^\prime)=Q(E,\eta|E^\prime,\eta^\prime)~
\end{align}
with $Q(E,\eta|E^\prime,\eta^\prime)$ obtained, as before~\cite{BF15,BF17a,BFP18,BF22}, 
from the invariant embedding principle~\cite{Dashen64,GP07,ARK20}.

\begin{figure}[t]
\includegraphics[width=0.99\linewidth]{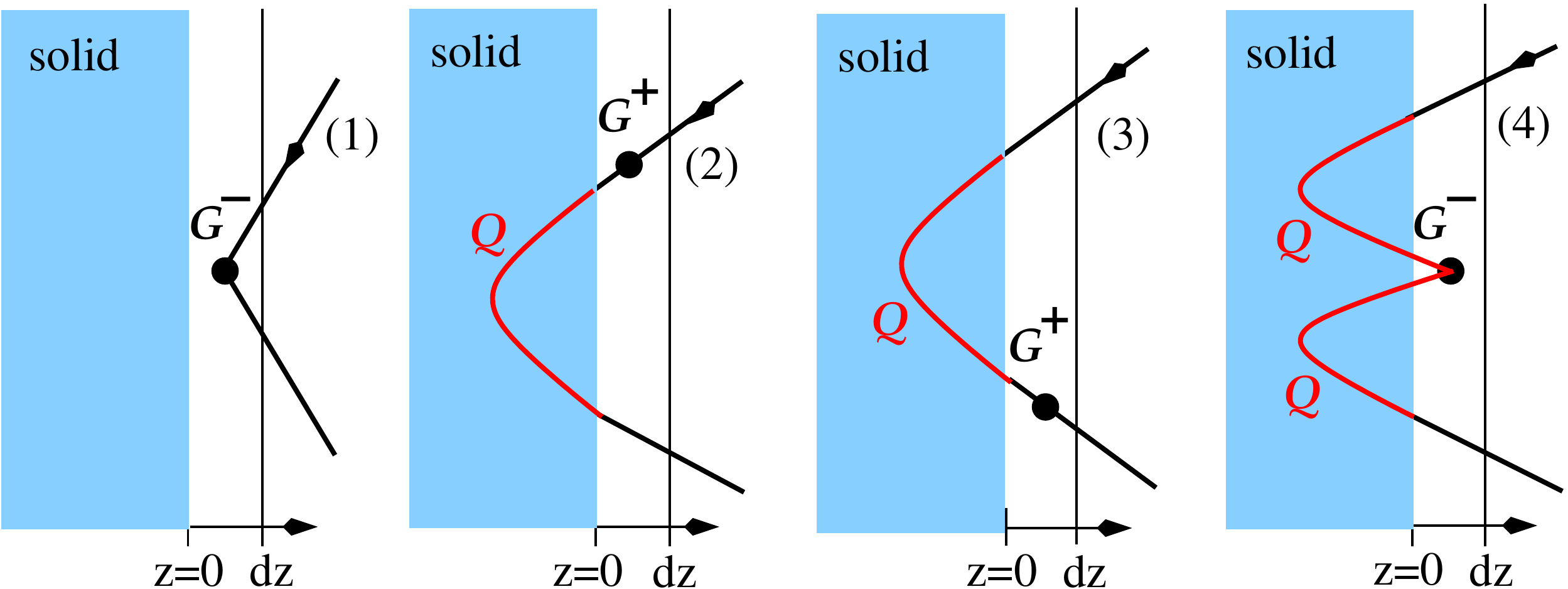}
\caption{(Color online) Illustration (adopted from Ref.~\cite{BF17a}) of the embedding principle 
leading to the nonlinear integral equation~\eqref{EmbeddingEq}. Due to the scattering in the 
infinitesimally thin layer on top of the halfspace filled with the solid, an electron hitting the 
interface has four additional pathways (1)--(4) available. However, the four paths should not change 
the backscattering function $Q(E,\eta|E^\prime,\eta^\prime)$. Hence, the increase in backscattering 
the four paths induce must cancel with the decrease of backscattering due to the old paths,
which is the original $Q(E,\eta|E^\prime,\eta^\prime)$ multiplied with the probability that  
scattering occurs at all on the inward and outward bound legs of the path through the 
layer~\cite{Dashen64}.
}
\label{EmbedCartoon}
\end{figure}

In its basic form, given by Dashen~\cite{Dashen64} and illustrated in Fig.~\ref{EmbedCartoon}, 
the principle states that adding an infinitesimally thin layer of the same material on top of a 
halfspace filled with it does not change the backscattering. Symmetrizing the 
backscattering function,
\begin{align}
Q(E,\eta|E^\prime,\eta^\prime)\rightarrow\sqrt{\rho(E)}Q(E,\eta|E^\prime,\eta^\prime)\sqrt{\rho(E^\prime)}~,
\label{Symmetrization}
\end{align}
the principle leads to the embedding equation,  
\begin{align}
G^- + (G^+ - S) \circ Q + Q \circ (G^+ - S) + Q \circ G^- \circ Q = 0~,
\label{EmbeddingEq}
\end{align}
where the $\circ$ operation is defined by 
\begin{align}
&(A \circ B)(E,\eta|E^\prime,\eta^\prime)=\nonumber\\
&\int_{E^\prime}^E \int_0^1
dE^{\prime\prime}d\eta^{\prime\prime}
A(E,\eta|E^{\prime\prime},\eta^{\prime\prime})
B(E^{\prime\prime},\eta^{\prime\prime}|E^\prime,\eta^\prime)~.
\label{CircDef}
\end{align}

The kernels, inverse scattering lengths weighted by the direction of propagation, can be
obtained from the Golden Rule transition rates $W^\pm(E,\eta|E^\prime,\eta^\prime)$ 
for forward $(+)$ and backward ($-$) scattering, 
\begin{align}
G^\pm(E,\eta|E^\prime,\eta^\prime)=\sqrt{\rho(E)}\,\frac{W^\pm(E,\eta|E^\prime,\eta^\prime)}{v(E)\eta}\sqrt{\rho(E^\prime)}~,
\label{Gfct}
\end{align}
whereas 
\begin{align}
S(E,\eta|E^\prime,\eta^\prime)=\frac{\Gamma(E)}{v(E)\eta}\delta(E-E^\prime)\delta(\eta-\eta^\prime)~ 
\end{align}
with 
\begin{align}
\Gamma(E)&=\int_{-\chi}^E \!\! dE^{\prime} \int_{0}^1 \!\! d\eta^{\prime}
\rho(E^{\prime})
\bigg[W^+(E,\eta|E^{\prime},\eta^{\prime})
\nonumber\\
&+ W^-(E,\eta|E^{\prime},\eta^{\prime})\bigg]~,
\label{GammaFct}
\end{align}
the total scattering rate at energy $E$ which in fact is independent of $\eta$. 
Within the bare electron mass model described in the previous subsection
the velocity of an electron with energy $E$ becomes $v(E)=2\sqrt{E+\chi}$ and the 
transition rates $W^\pm(E,\eta|E^\prime,\eta^\prime)$ will be specified in the next 
subsection. The numerical strategy to solve the embedding equation~\eqref{EmbeddingEq} 
follows Shimizu and coworkers~\cite{SM66a,SA72}
who used the invariant embedding approach to study the shielding of $\gamma-$rays in nuclear 
reactors. It is sketched in Appendix~\ref{Numerics} and solves the equation without 
approximation except the discretization of the integrals.\\

\subsection{Randium-jellium model}
\label{Model}

In the previous subsection we described a general scheme for the construction of the surface 
scattering kernel $R(E,\xi|E^\prime,\xi^\prime)$. To compute numerical values, we have to 
furnish the approach with a microscopic model for the solid, that is, we have to specify 
the electronic structure, the interface potential, and the scattering processes inside the
semiconductor. Having in mind applying the model to other materials as well,
we keep it as generic as possible.

Inspired by the work of Bauer and coworkers~\cite{Bauer70,DMB82}, we employ a 
randium-jellium-type model, where the ion cores of the solid 
are randomly immersed in an electron liquid. The elastic scattering of electrons 
on the ion cores is assumed to be incoherent and 
described by a screened pseudopotential which is also used in electronic band structure 
calculations~\cite{ILC78,SCL75,Phillips73}. Screening is subtle in covalently bound solids.
We account for it phenomenologically along the lines of Penn~\cite{Penn62}, augmented by ideas 
of Phillips~\cite{Phillips68} and parameters of Srinivasan~\cite{Srinivasan69}. In addition
to the scattering on the ion cores, we consider electron-phonon scattering, and as the main
energy loss process, impact ionization across the energy gap. The latter causes also 
electron multiplication and is thus of central importance for secondary electron emission. 

\begin{table}[t]
\begin{center}
  \begin{tabular}{c|c|c|c|c|c|c|c}
 \hline\hline
   & $Z$ & $a[\AA]$ & $\chi[\mathrm{eV}]$ & $E_{\rm g}[\mathrm{eV}]$ & $\omega_{\rm LO}[\mathrm{eV}]$ & $D_tK[10^8\frac{{\rm eV}}{{\rm cm}}]$ & $\varepsilon$ \\\hline\hline
Si & 4   & 5.43     & 4.05                & 1.11                     & 0.063                               & 11  & 11.7    \\
Ge & 4   & 5.66     & 4.0                 & 0.66                     & 0.037                               & 9.5 & 16.2    \\\hline\hline
   &     &          &                     &                          &                                     &     &         \\\hline\hline
         & $a_1$    & $a_2$ & $a_3$ & $a_4$ & $E_{\rm g}^{\rm ave}[\mathrm{eV}]$ & $\rho[{\rm g}/{\rm cm}^3]$ & $n_{\rm ion}$ \\\hline\hline
Si & -0.992 & 0.791 & -0.352  & -0.018  &  4.8  & 2.33  & 0.0073        \\
Ge & -0.955 & 0.803 & -0.312  & -0.019  &  4.2  & 5.32  & 0.0065        \\\hline\hline
  \end{tabular}
  \caption{\small Valence $Z$, lattice constant $a$, electron affinity $\chi$, energy gap $E_{\rm g}$,
           phonon energy $\omega_{\rm LO}$, optical deformation potential $D_tK$, dielectric
           constant $\varepsilon$, pseudopotential parameters~\cite{ILC78,SCL75} $a_i$,
           average optical energy gap~\cite{Penn62,Srinivasan69} $E_{\rm g}^{\rm ave}$,
           mass density $\rho$, and atomic density $n_{\rm ion}$. If not noted otherwise, the 
           material parameters are from Jacoboni and Reggiani~\cite{JR83} and given in atomic
           units, with energy measured in Rydbergs, length in Bohr radii, and mass in bare 
           electron masses.
  }
  \label{MaterialParameters}
\end{center}
\end{table}

The full band structure of the solid cannot be represented by the randium-jellium model.
We hence approximate the electronic structure of the semiconductor by a parabolic conduction
and a parabolic valence band, separated by a direct energy gap $E_g$, and both with effective
mass equal to the bare electron mass (see also discussion preceding 
Eq.~\eqref{DeltaR}). Taking then a Schottky barrier~\cite{MacColl39} with 
depth $\chi$ as an interface potential, as indicated in Fig.~\ref{ModelCartoon}c, the probability 
for an electron to be quantum-mechanically reflected from the interface region reads~\cite{RBF22}
\begin{align}
R(E,\xi)=\Bigg|\frac{\sqrt{\widetilde{E}_z}-\sqrt{E_z}y}{\sqrt{\widetilde{E}_z}+\sqrt{E_z}y^*}\Bigg|^2
\end{align}
with $E_z=E\xi^2$, $\widetilde{E}_z=E_z+\chi$, and 
\begin{align}
y=-2\,\frac{W^\prime_{\lambda,1/2}(\xi_0)}{W_{\lambda,1/2}(\xi_0)}~,
\end{align}
where $W_{\lambda,1/2}(x)$ is a Whittaker function, $W^\prime_{\lambda,1/2}(x)$ its derivative 
with respect to its argument,   
\begin{align}
\lambda &= -i\frac{\varepsilon-1}{\varepsilon+1}\frac{1}{\sqrt{8E_z}}~,\\
\xi_0 &=i\frac{\sqrt{2}}{\chi}\frac{\varepsilon-1}{\varepsilon+1}\sqrt{E_z}~, 
\end{align}
$y^*$ is the complex conjugate of $y$, and $\varepsilon$ is the dielectric constant of the 
solid. As in the work for metals~\cite{BF22}, energy gaps in the reflection probability could
be included. But the experimental data for the emission yield of silicon and 
germanium~\cite{FF58,BronFrai69}, the materials we use as an illustration of our approach,
do not indicate that this is required.

We are now turning to the scattering processes inside the solid. Measured from the 
conduction band minimum, electron emission and reflection take place at energies much 
larger than phonon energies, even the energy of the longitudinal optical phonon 
$\omega_{\rm LO}\ll E$. It is thus appropriate to describe electron-phonon scattering 
quasi-elastically and to combine it with electron-ion-core scattering to a single 
elastic scattering process. Since the latter
should be absent at low energies, close to the band minimum, we adopt an idea of Kieft 
and Bosch~\cite{KB08} and switch linearly between two threshold energies from 
electron-phonon to electron-ion-core scattering. Below the lower threshold,
$E^{\rm th}_1=(2\pi/a)^2-\chi$, taken to be the energy measured from the bottom of 
the conduction band for which the de Broglie wavelength of the electron is equal to the 
lattice constant $a$, elastic scattering is due to phonons, whereas above the upper 
threshold, $E^{\rm th}_2=3\,(2\pi/a)^2-\chi$, elastic scattering is due to ion cores. 
The upper threshold is chosen in such a way to get good agreement with measured 
emission yields. Albeit ad-hoc, it seems plausible to assume 
that electrons with small wave length do not notice the lattice 
periodicity, especially when they propagate in arbitrary directions. Hence,
they suffer mostly binary collisions with the ion cores and not scattering on collective 
modes of the crystal lattice. Indeed, at energies 
above 100~eV neglecting the periodicity of the lattice potential seems to be generally 
accepted (but see the recent discussion by Werner~\cite{Werner23}).
Systematic work, based on quantum-kinetic equations, which so far has been only done 
for high energies~\cite{DPW93}, is however required to ultimately clarify this point. 

\begin{figure}[t]
\begin{minipage}{0.5\linewidth}
\rotatebox{270}{\includegraphics[width=0.85\linewidth]{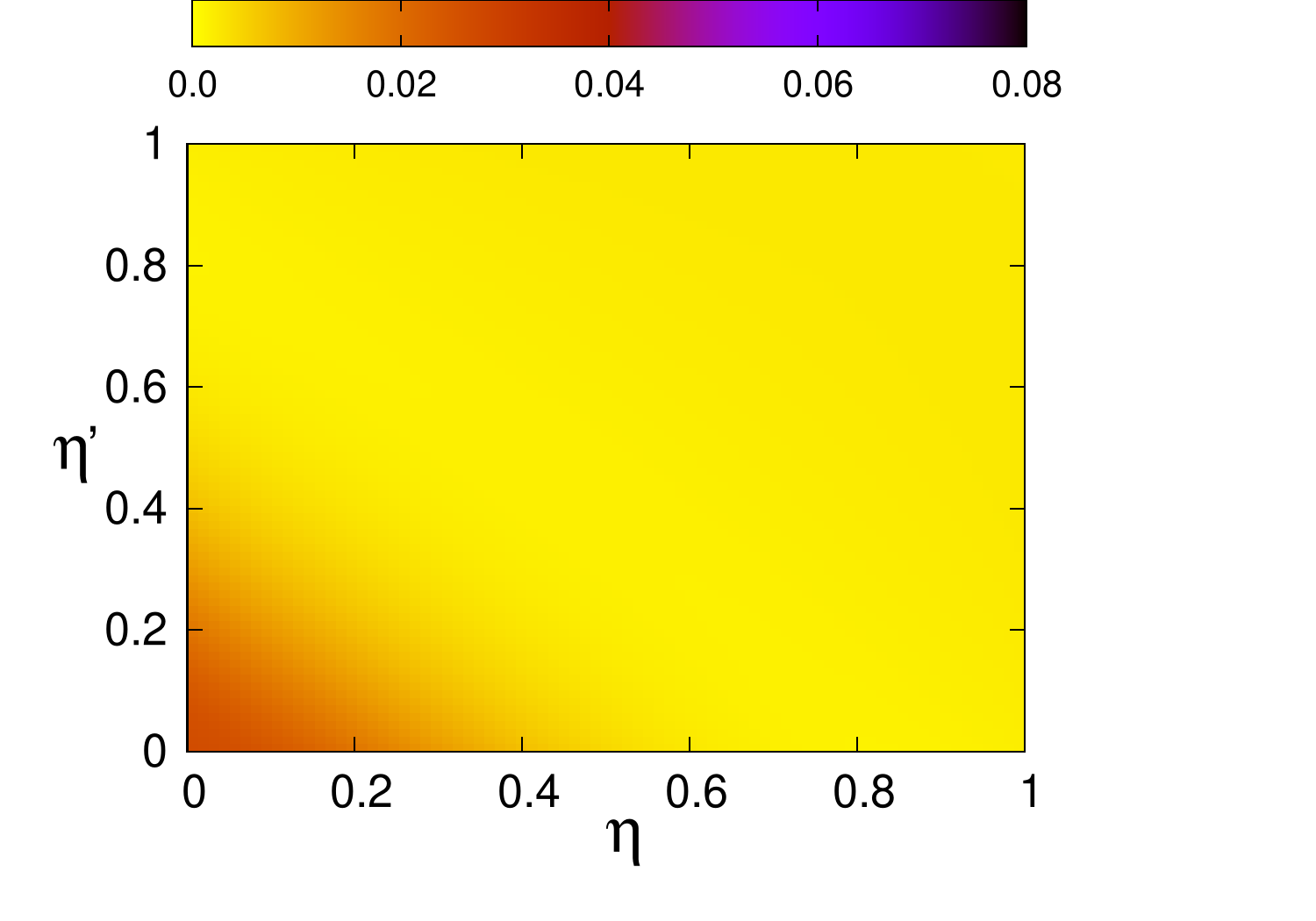}}

\rotatebox{270}{\includegraphics[width=0.85\linewidth]{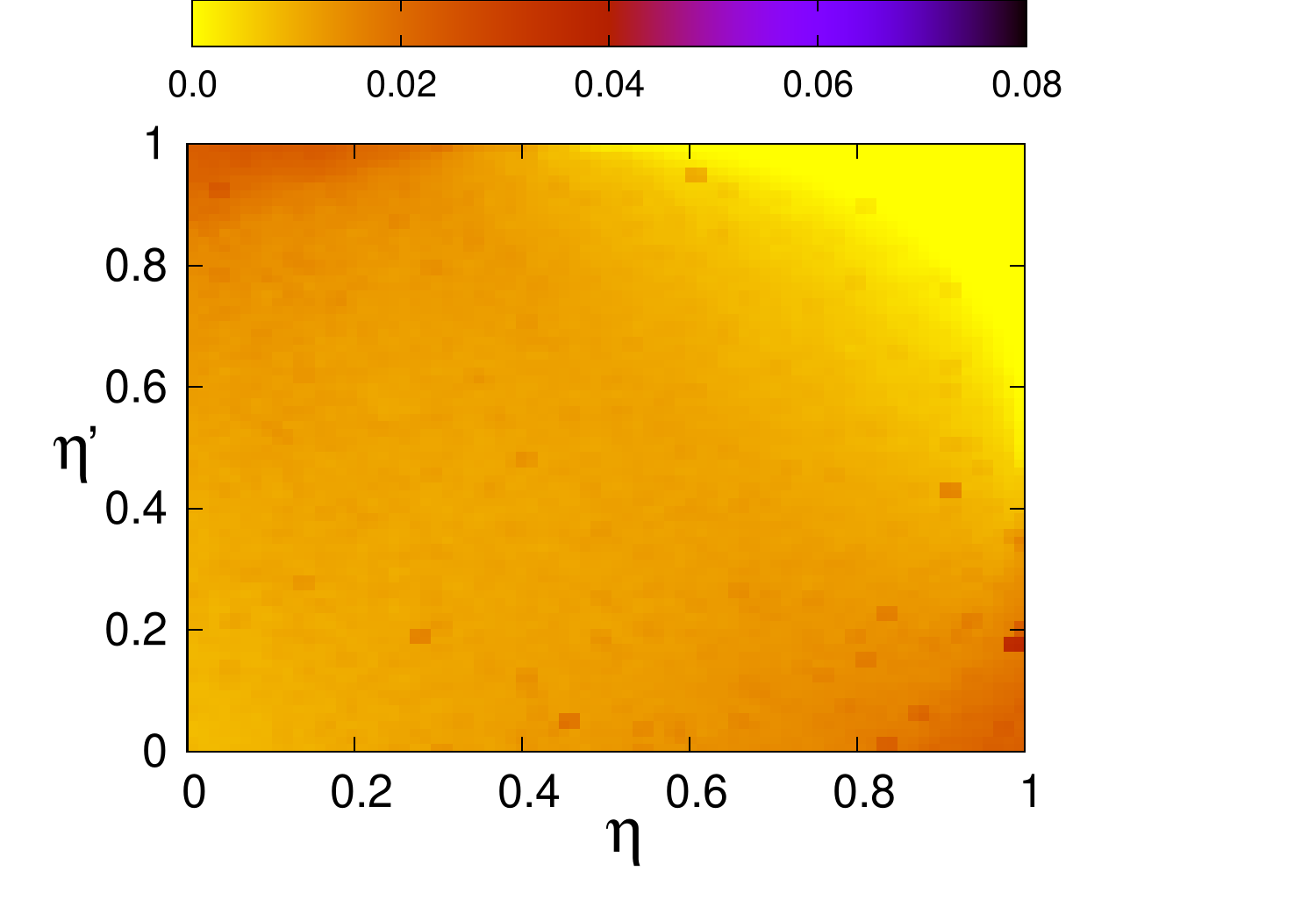}}

\end{minipage}\begin{minipage}{0.5\linewidth}
\rotatebox{270}{\includegraphics[width=0.85\linewidth]{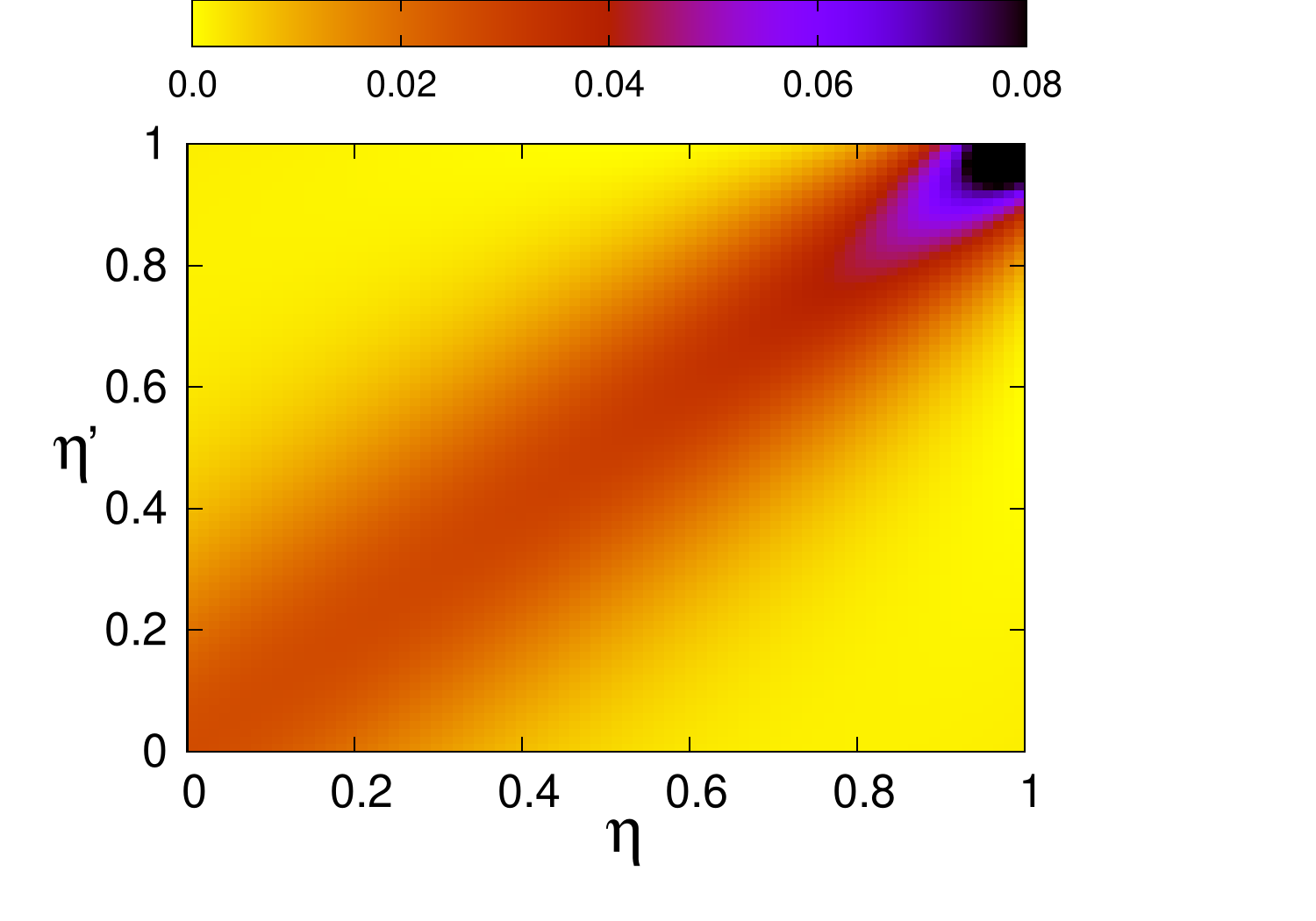}}

\rotatebox{270}{\includegraphics[width=0.85\linewidth]{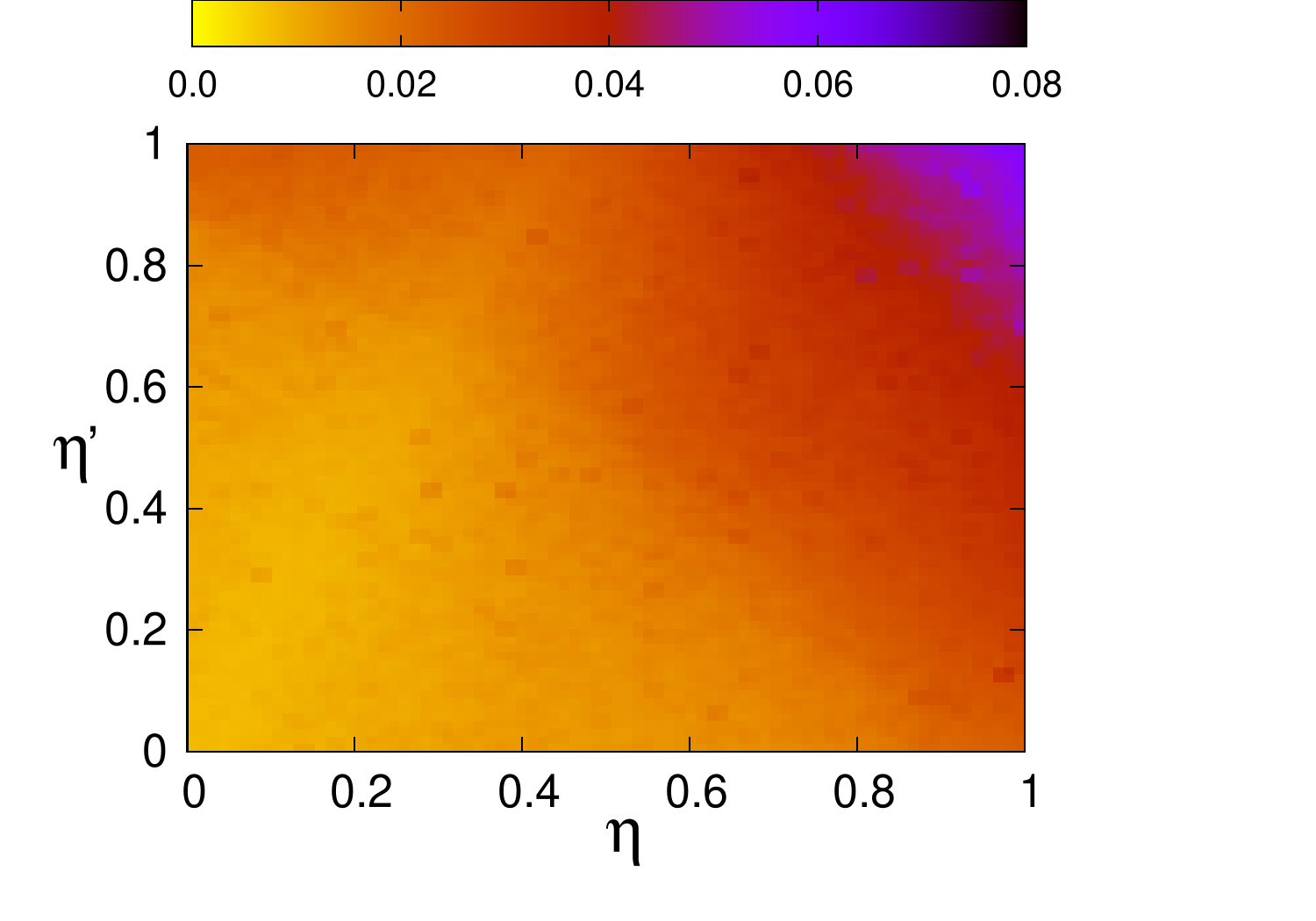}}

\end{minipage}
\caption{(Color online) Angle dependence of the transition rates for Si using
the model described in~\ref{Model}. Left and right panels display for initial
energy $E=28.8\,{\rm eV}$ and, from top to bottom, final energies
$E^\prime=28.8\,{\rm eV}\, {\rm and}\, 1.8\,{\rm eV}$ the
rates $W^-(E,\eta|E^\prime,\eta^\prime)$ and $W^+(E,\eta|E^\prime,\eta^\prime)$
as employed for the calculation of $\Gamma(E)$, that is, without the factor two
in front of $W^\pm_{\rm impact}(E,\eta|E^\prime,\eta^\prime)$.
The noise of the Monte Carlo integration required for the computation of the
impact ionization transition rate is clearly seen in the lower two panels.
}
\label{RatesAngleDependence}
\end{figure}

The starting point for the calculation of the elastic transition rates is the standard
Golden Rule expression. Introducing
spherical coordinates in momentum space, with the $z-$axis pointing along the inward 
interface normal, states with $k_z>0$ describe electrons moving inwards the solid, whereas 
states with $k_z<0$ denote states moving outwards. It is then straightforward to work out 
the rates for forward $(+)$ and backward $(-)$ scattering and to express them in terms 
of the variables defined in Fig.~\ref{ModelCartoon}b: the total energy $E$, the direction 
cosine $\eta$, and the azimuth angle $\Phi$, which for a laterally homogeneous interface 
does however not appear because it is integrated out. As a result, one obtains 
\begin{align}
W_{\rm elastic}^\pm(E,\eta|E^\prime,\eta^\prime)&=W_{\rm ep}^\pm(E,\eta|E^\prime,\eta^\prime)
	\Theta_{1}(E, E^{\rm th}_1, E^{\rm th}_2)\nonumber\\
&+W_{\rm eic}^\pm(E,\eta|E^\prime,\eta^\prime)\Theta_2(E, E^{\rm th}_1, E^{\rm th}_2)
\end{align}
with $\Theta_{1,2}(E, E^{\rm th}_1, E^{\rm th}_2)$ auxiliary functions enforcing the 
linear switching  described above, 
\begin{align}
W_{\rm ep}^\pm(E,\eta|E^\prime,\eta^\prime)=\frac{M^2}{2\pi}[1+2n_B(\omega_{\rm LO})]\delta(E-E^\prime)~
\label{Wep}
\end{align}
the rate for electron-(longitudinal optical) phonon scattering, where 
$M^2=(D_tK)^2/(\omega_{\rm LO}\rho)$ is the scattering strength and $n_B(\omega)=1/(\exp(\beta\omega)-1)$
the Bose function, and 
\begin{align}
W_{\rm eic}^\pm(E,\eta|E^\prime,\eta^\prime)=\frac{1}{(2\pi)^2n_{\rm ion}}
\langle |U_{\rm ps}(g^\pm)|^2\rangle_\Phi \delta(E-E^\prime)~
\label{Weic}
\end{align}
the electron-ion-core scattering rate with $\langle ... \rangle_\Phi=\int_0^{2\pi}(...)\mathrm{d}\Phi$ 
denoting the integral over the azimuth angle, $n_{\rm ion}$ the atomic density of the solid, and 
\begin{align}
U_{\rm ps}(q)=\frac{Z/\bar{\varepsilon}}{q^2+k_s^2}a_1(\cos(a_2q)+a_3)\exp(a_4 q^4) 
\label{PseudoPotential}
\end{align}
the Fourier transform of the pseudopotential~\cite{ILC78,SCL75,Phillips73} of an ion with 
valence $Z$, phenomenologically screened~\cite{Srinivasan69,Phillips68,Penn62} as explained 
in Appendix~\ref{Screening}. Its normalization leads to the factor $1/n_{\rm ion}$ in the 
scattering rate. Finally, 
\begin{align}
g^\pm &= |\vec{k}-\vec{k}^\prime|^\pm\nonumber\\
      &= g(E,T,p=1|E^\prime,T^\prime,p^\prime=\pm 1;\Phi)~,
\end{align}
where the function $g(E,T,p|E^\prime,T^\prime,p^\prime;\Phi)$ is defined in~\eqref{gtilde} 
and $T=(E+\chi)(1-\eta^2)$ and $T^\prime=(E^\prime+\chi)(1-(\eta^\prime)^2)$ are the lateral 
kinetic energies of the initial and final state. The material parameters entering the rates are 
named and listed in Table~\ref{MaterialParameters}.

\begin{figure}[t]
\begin{minipage}{0.5\linewidth}
\rotatebox{270}{\includegraphics[width=0.85\linewidth]{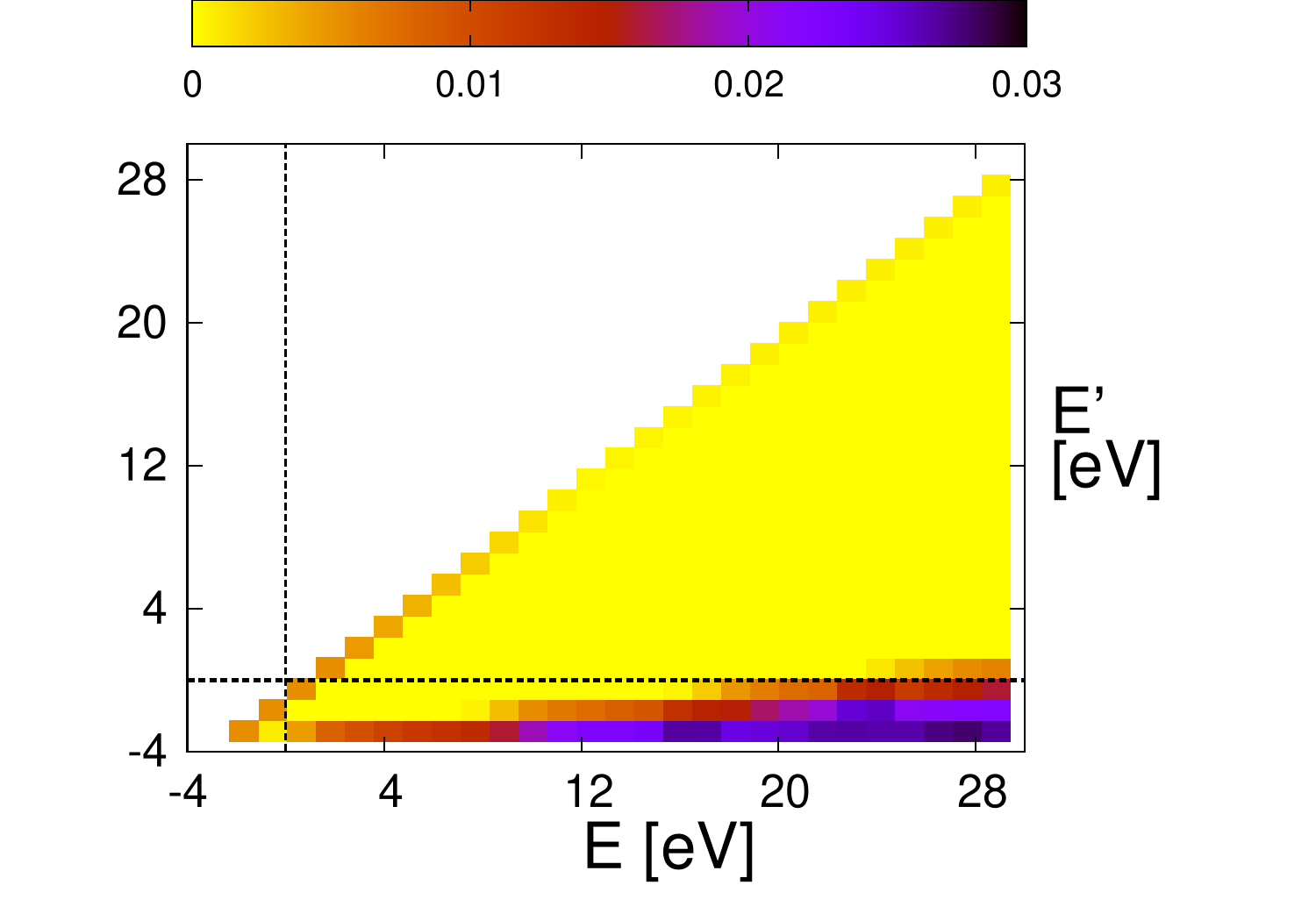}}

\rotatebox{270}{\includegraphics[width=0.859\linewidth]{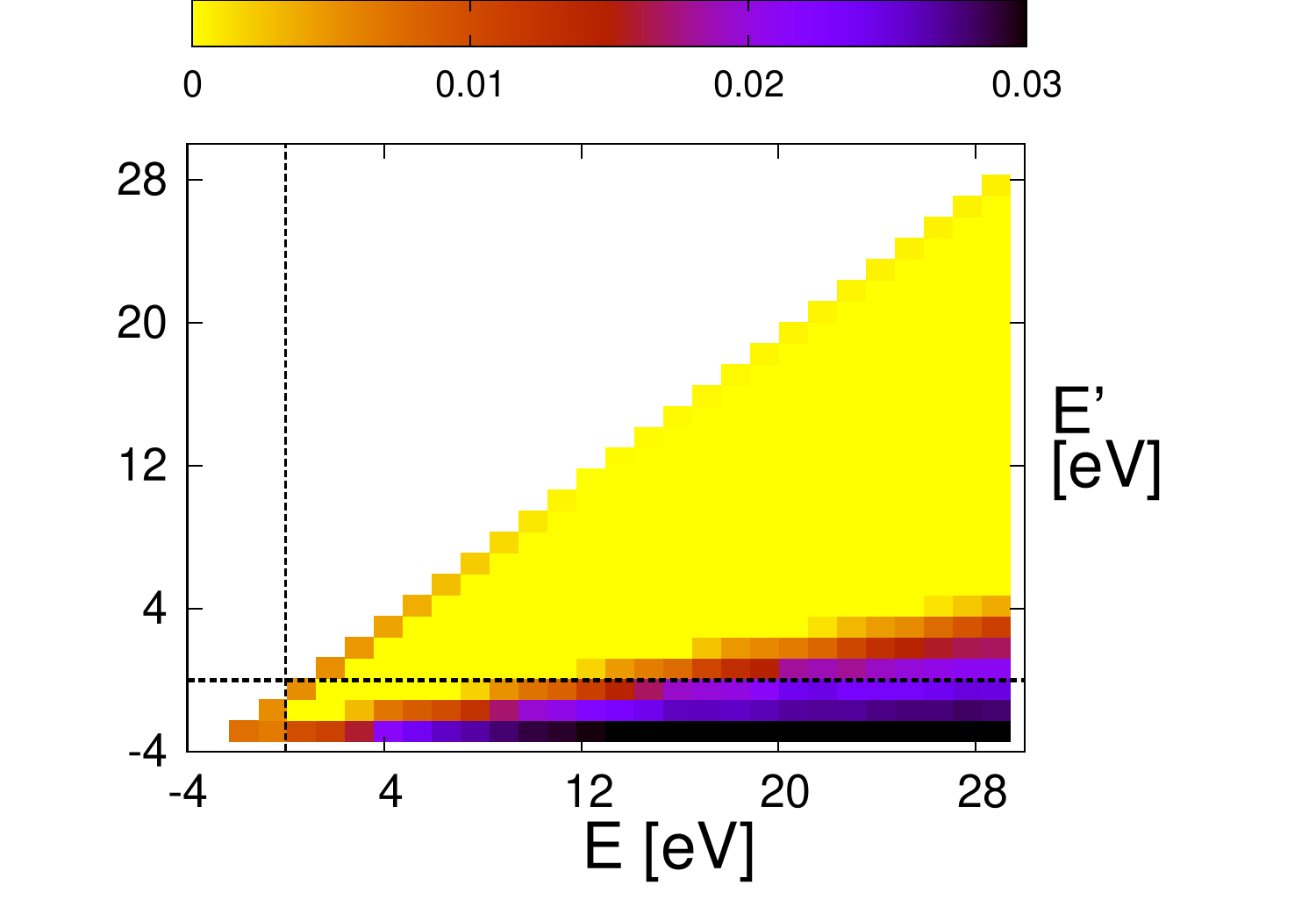}}

\end{minipage}\begin{minipage}{0.5\linewidth}
\rotatebox{270}{\includegraphics[width=0.85\linewidth]{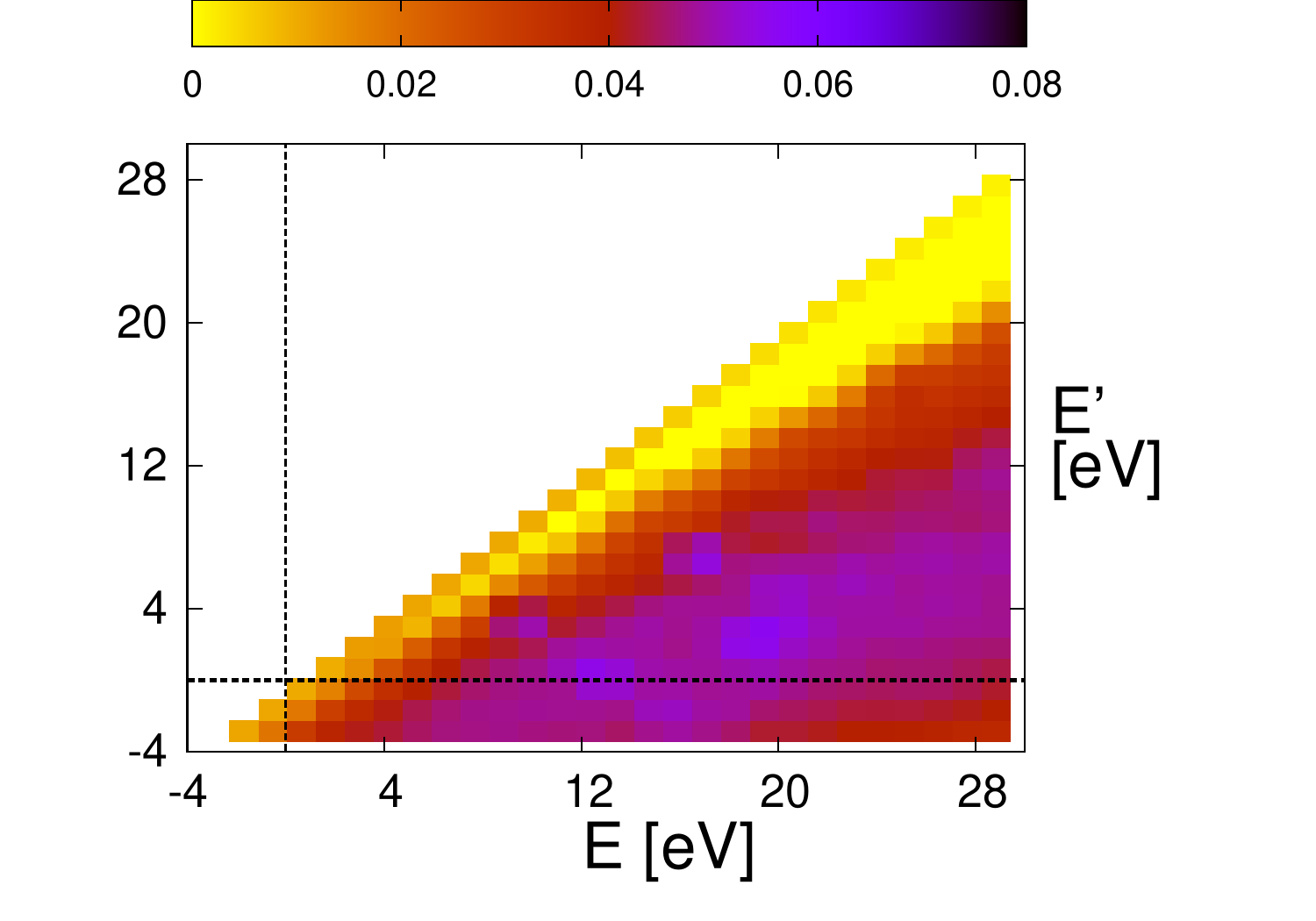}}

\rotatebox{270}{\includegraphics[width=0.85\linewidth]{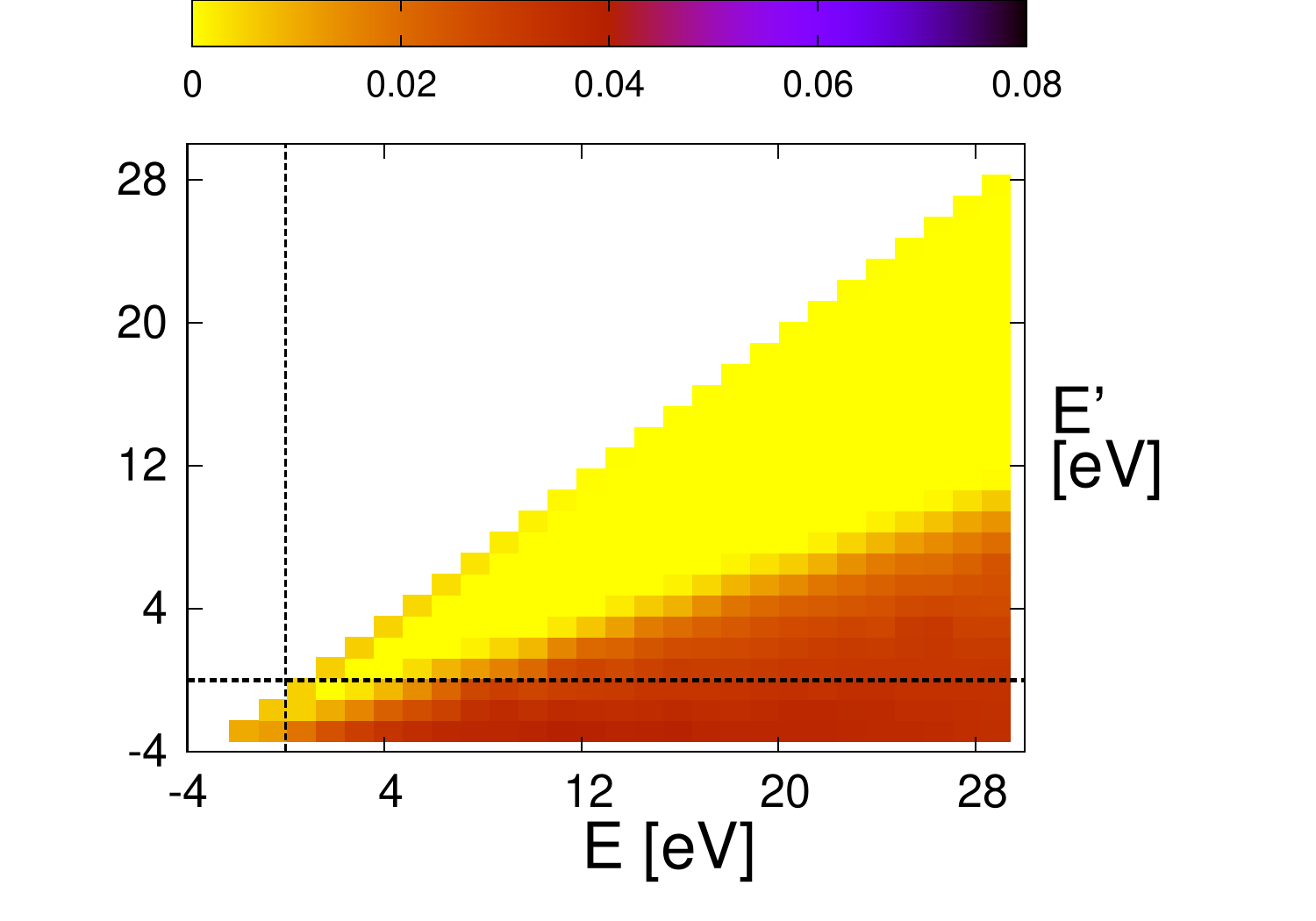}}

\end{minipage}
\caption{(Color online) Energy dependence of the transition rates for Si
using the model described in~\ref{Model}. Left and right panels
display $W^-(E,\eta|E^\prime,\eta^\prime)$ and $W^+(E,\eta|E^\prime,\eta^\prime)$
as employed for the calculation of $\Gamma(E)$, that is, without the factor
two in front of $W^\pm_{\rm impact}(E,\eta|E^\prime,\eta^\prime)$
for $\eta=1$ and, from top to bottom, $\eta^\prime=0.75$ {\rm and}\, $0.24$.
The dashed lines indicate the just-outside potential energy, which is set to zero.
}
\label{RatesEnergyDependence}
\end{figure}

For energies larger than the band gap, impact ionization is possible and becomes the main 
inelastic scattering process. To derive its rate, we switch to the hole representation 
for the valence band and start with the standard Golden Rule representation as given, for 
instance, by Kane~\cite{Kane67}. Using spherical coordinates in momentum space with the 
$z-$axis pointing again inwards, identifying scattering between states for forward and 
backward moving electrons, and employing the total energy $E$, the lateral kinetic energy 
$T$ (instead of the direction cosine $\eta$), and the azimuth angle 
$\Phi$ as independent variables, we obtain 
\begin{align}
W^\pm_{\rm impact}&(E,\eta|E^\prime,\eta^\prime)= 
{\cal W}(E,T,1|E^\prime,T^\prime,\pm 1)
\label{Wimpact}
\end{align}
with the function ${\cal W}(E,T,p|E^\prime,T^\prime,p^\prime)$ derived 
in Appendix~\ref{ImpactRate} and $T$ and $T^\prime$ the lateral kinetic energies 
given above. 

As illustrated in Fig.~\ref{ModelCartoon}c, impact ionization leads to two conduction band 
electrons~\cite{Kane67,TPE91,BHI92,CFE93}. To implement in our formalism the correct 
ionization rate, we thus have to avoid double counting by correctly normalizing, respectively, 
the contribution of impact ionization to the kernels $G^\pm(E,\eta|E^\prime,\eta^\prime)$ 
and to the total scattering rate $\Gamma(E)$. Following the reasoning of Penn and coworkers~\cite{PAG85}, 
as well as  Wolff's~\cite{Wolff54}, we find $W^\pm_{\rm impact}(E,\eta|E^\prime,\eta^\prime)$ 
to be multiplied by a factor two if used in~\eqref{Gfct} for the kernels 
$G^\pm(E,\eta|E^\prime,\eta^\prime)$, whereas the plain $W^\pm_{\rm impact}(E,\eta|E^\prime,\eta^\prime)$ 
enters~\eqref{GammaFct} for the total scattering rate $\Gamma(E)$. Hence, in total, the transition 
rates to be used in the kernels $G^\pm(E,\eta|E^\prime,\eta^\prime)$ read 
$W^\pm(E,\eta|E^\prime,\eta^\prime)=W^\pm_{\rm elastic}(E,\eta|E^\prime,\eta^\prime)
+2\,W^\pm_{\rm impact}(E,\eta|E^\prime,\eta^\prime)$, while in the total 
scattering rate $\Gamma(E)$ the rates   
$W^\pm(E,\eta|E^\prime,\eta^\prime)=W^\pm_{\rm elastic}(E,\eta|E^\prime,\eta^\prime)
+W^\pm_{\rm impact}(E,\eta|E^\prime,\eta^\prime)$ have to be inserted. 

\section{Results}
\label{Results}

\begin{figure}[t]
\includegraphics[width=0.99\linewidth]{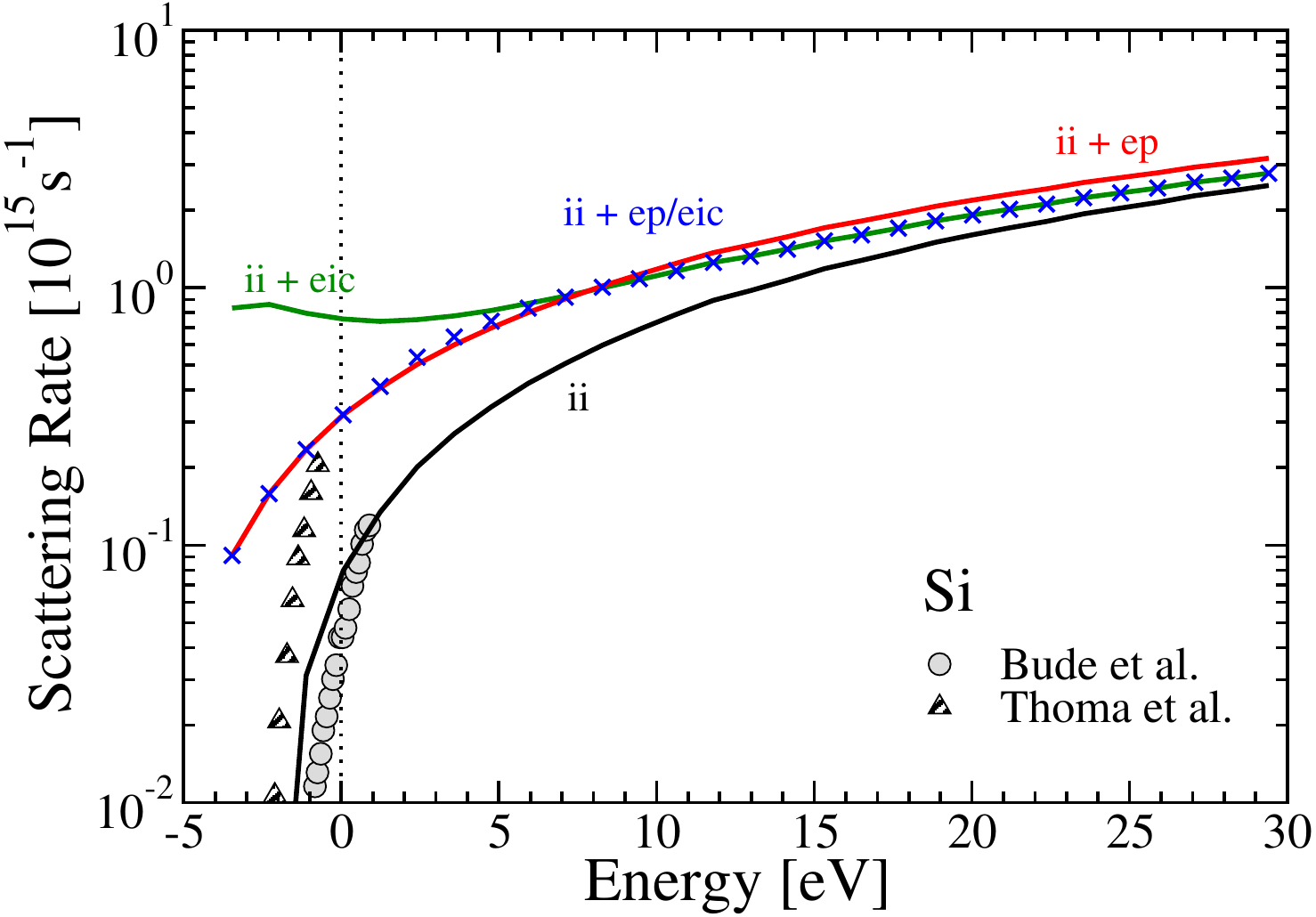}
\caption{(Color online) Total scattering rate $\Gamma(E)$ defined in~\eqref{GammaFct} 
for silicon using the material parameters of Table~\ref{MaterialParameters}.
For impact ionization alone (ii), $\Gamma(E)$ is in reasonable agreement
with results obtained by Bude~\cite{BHI92} and
Thoma~\cite{TPE91} et al. (as extracted from Fig.~1 of
Cartier and coworkers~\cite{CFE93}). Adding on top of it also
electron-phonon (ii+ep), electron-ion-core (ii+eic), or both (ii+ep/eic),
with the smooth linear switching between the two discussed in~\ref{Model} and 
visible by the blue crosses, $\Gamma(E)$ changes as shown. 
}
\label{GammaFig}
\end{figure}

Having established a model for the semiconducting interface, from which the 
transition rates $W^\pm(E,\eta|E^\prime,\eta^\prime)$ follow, the kernels 
of the embedding equation are given and we can solve~\eqref{EmbeddingEq}
by the numerical strategy explained in Appendix~\ref{Numerics} to obtain--at 
the end--the surface scattering kernel $R(E,\xi|E^\prime,\xi^\prime)$ as 
well as the emission yield $Y(E,\xi)$. To demonstrate the feasibility of 
our approach we consider silicon and germanium. The parameters of the 
model are given in Table~\ref{MaterialParameters} and the temperature of 
the solid is $300\,{\rm K}$. We are particularly interested in low electron
impact energies. Setting $E_{\rm max}=30\,{\rm eV}$ gives a width 
$\Delta E\simeq 1.2\,{\rm eV}$ for the $N=30$ energy windows into which we 
split in the numerical implementation the whole energy range from $-\chi$ to 
$E_{\rm max}$. The $M=80$ discretization steps of the integrals over the 
direction cosines lead to $\Delta\eta\simeq 0.0127$. 

\subsection{Validation of the semiconductor model}

\begin{figure}[t]
\includegraphics[width=0.99\linewidth]{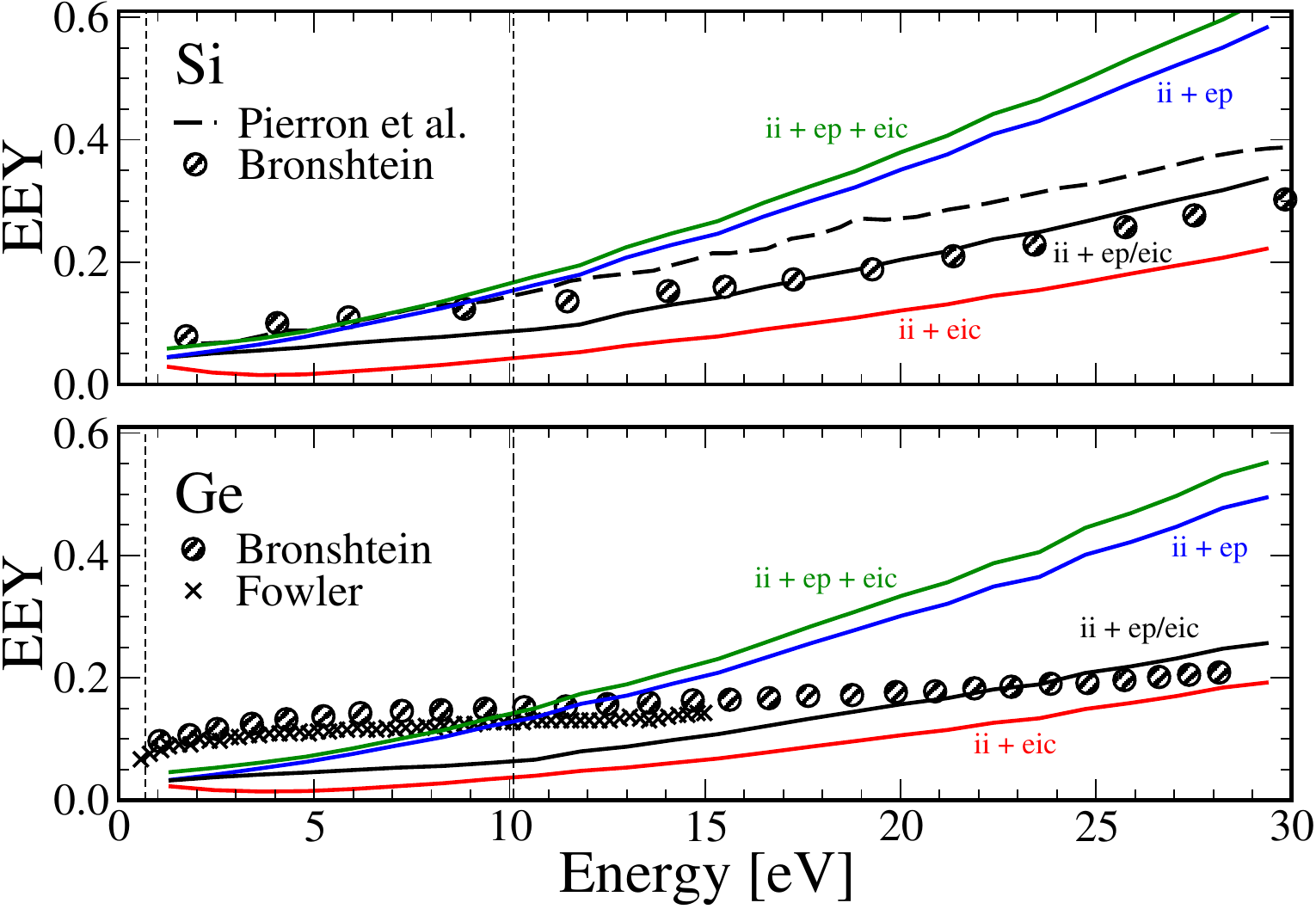}
\caption{(Color online) Secondary electron emission yield $Y(E,\xi)$ for a silicon and
germanium surface after it was hit perpendicularly ($\xi=1$) by an electron
with energy $E$. Experimental data are from Bronshtein and Fraiman~\cite{BronFrai69}
as well as Fowler and Farnsworth~\cite{FF58}. Monte Carlo data are from Pierron
and coworkers~\cite{PIB17}. Theoretical results are shown for various elastic 
scattering processes on top of impact ionization: electron-phonon and 
electron-ion-core scattering throughout the whole energy range (ii+ep+eic), 
electron-phonon scattering throughout the whole energy range (ii+ep), scattering on
ion cores throughout the whole energy range (ii+eic), and smooth linear switching 
between electron-phonon and electron-ion-core scattering as discussed in the main 
text (ii+ep/eic). The vertical dashed lines indicate the threshold energies
$E^{\rm th}_1$ and $E^{\rm th}_2$ between which the switching occurs.}
\label{YieldFig}
\end{figure}

\begin{figure*}[t]
\begin{minipage}{0.333\linewidth}
\includegraphics[width=0.98\linewidth]{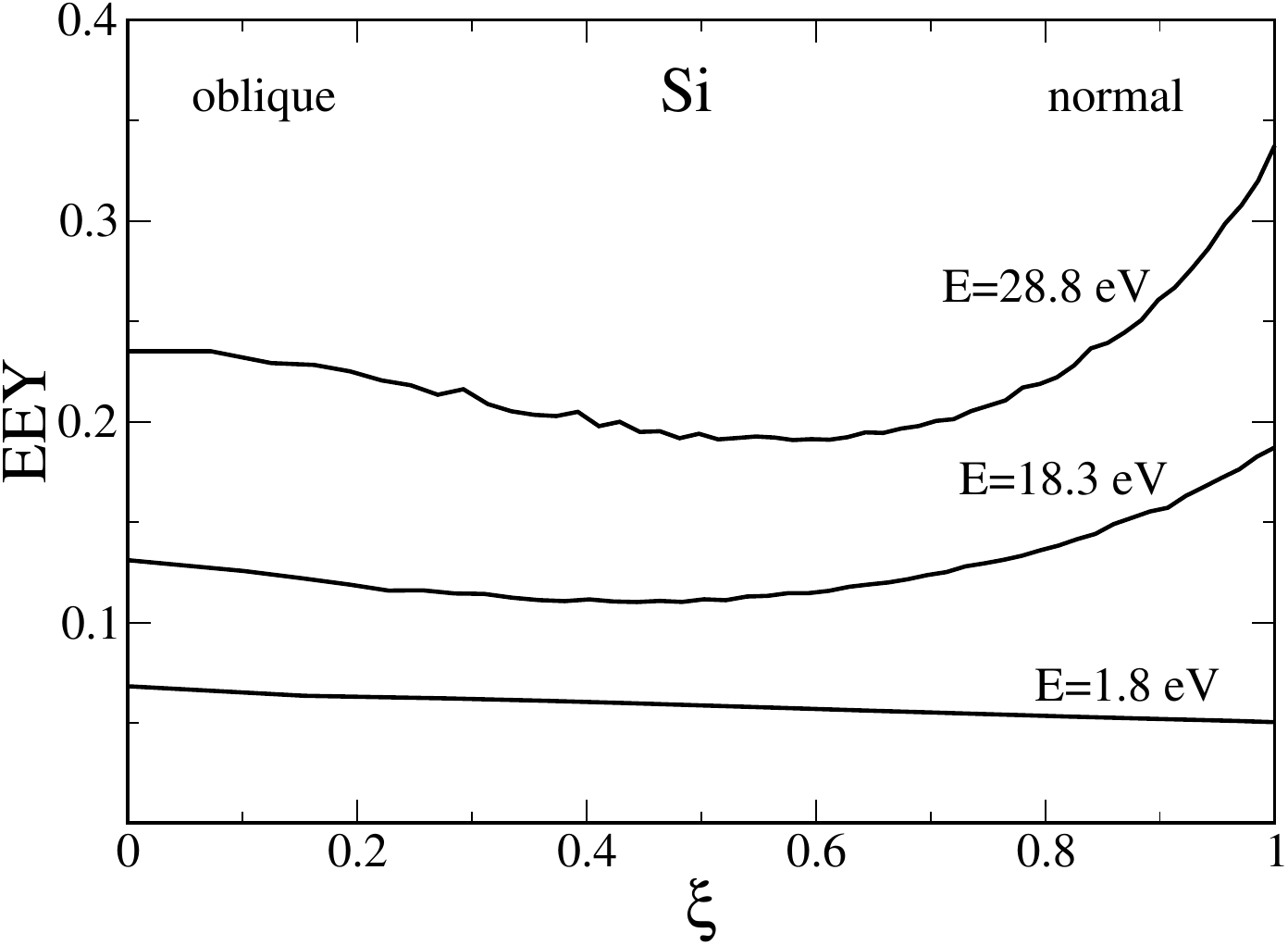}
\end{minipage}\begin{minipage}{0.333\linewidth}
\center{\rotatebox{270}{\includegraphics[width=0.85\linewidth]{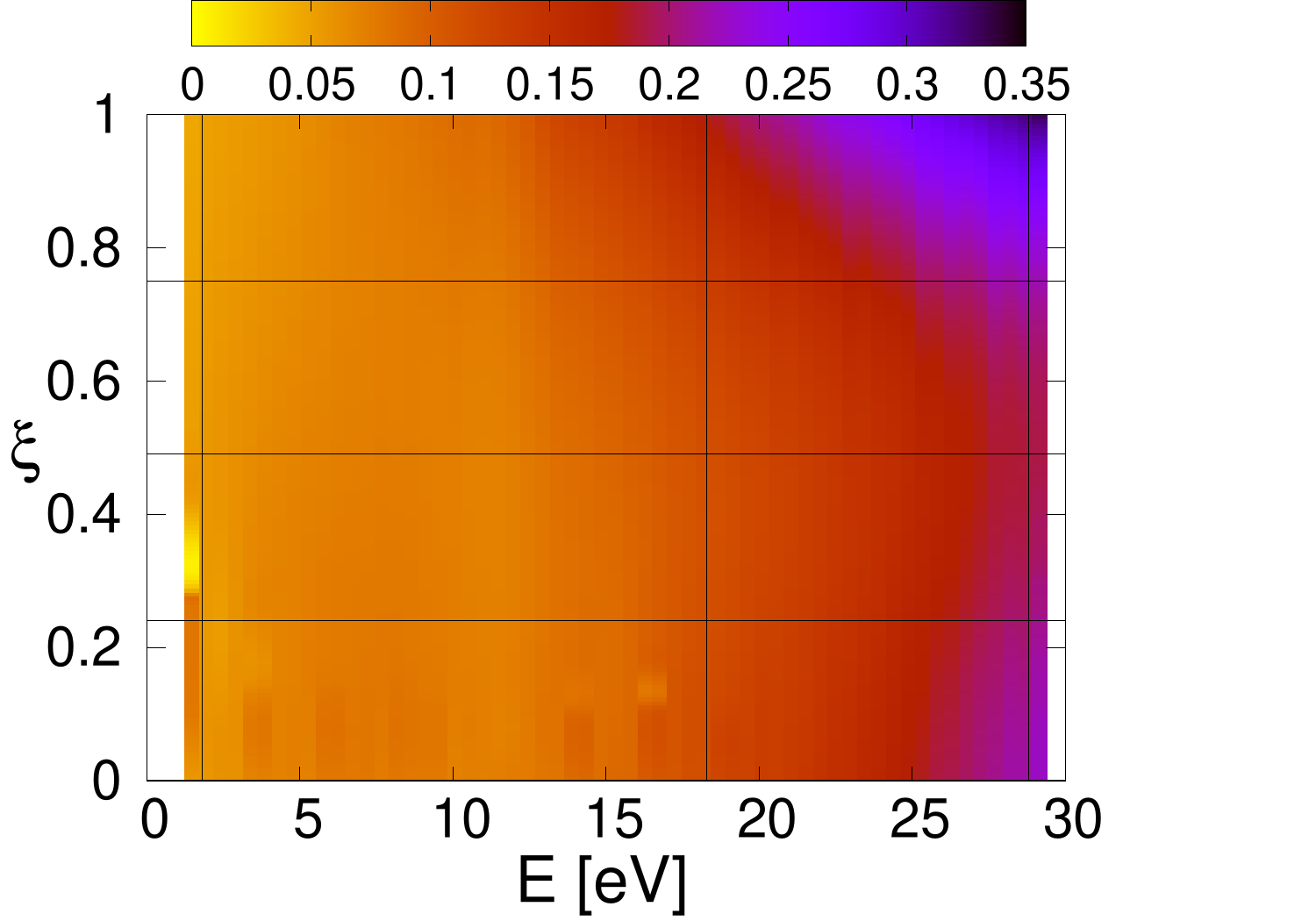}}}
\end{minipage}\begin{minipage}{0.333\linewidth}
\hfill\includegraphics[width=0.98\linewidth]{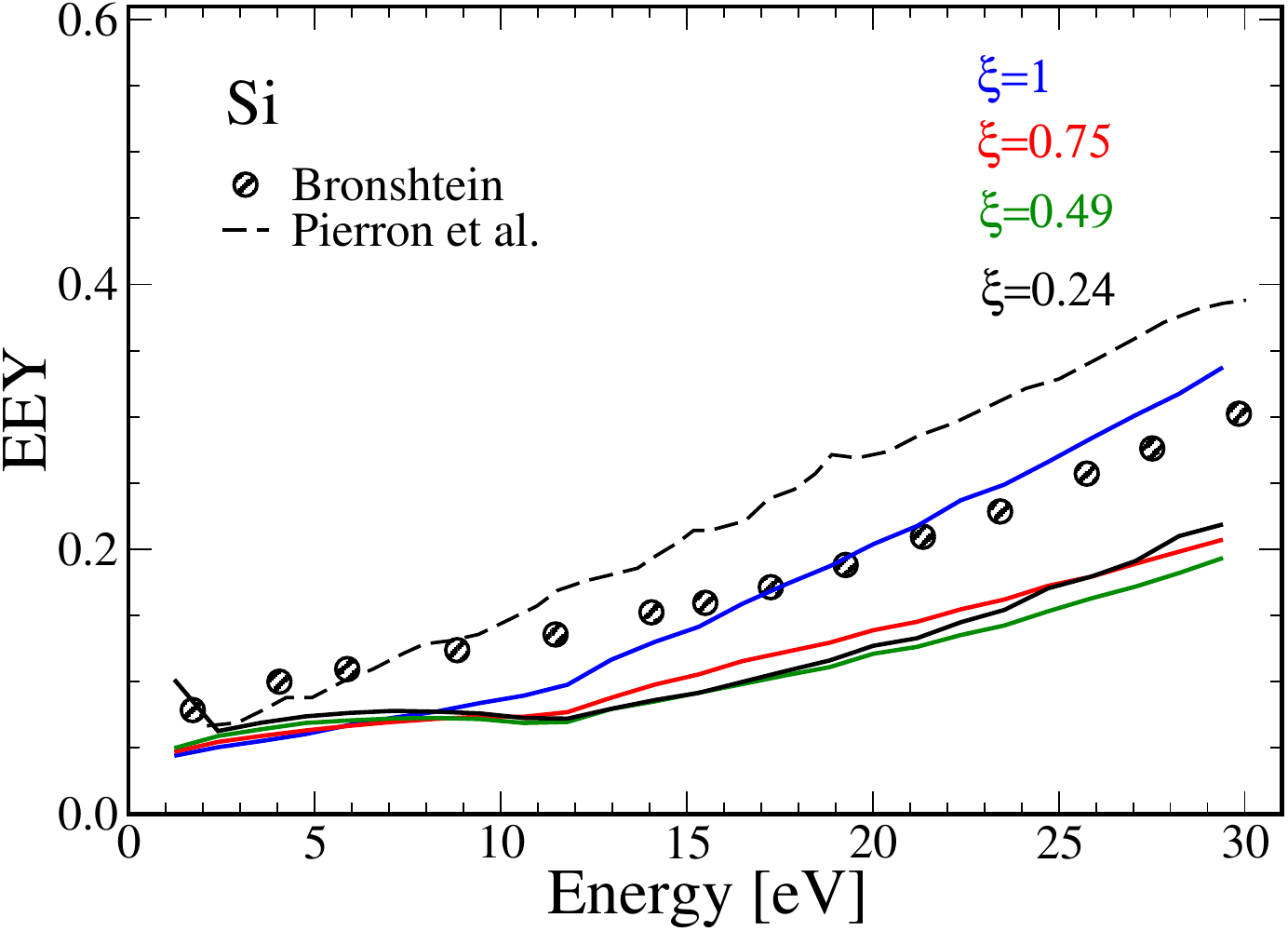}
\end{minipage}
\caption{(Color online) Center: Secondary electron emission yield $Y(E,\xi)$ over the whole 
range of impact energies $E$ and direction cosines $\xi$ considered in this work. Left: Angle 
dependence of the yield for the three impact energies indicated by the thin vertical lines: 
$E=1.8\,\mathrm{eV}$, $18.3\,\mathrm{eV}$, and $28.8\,\mathrm{eV}$. Right: Energy dependence 
of the yield for the three direction cosines indicated by the thin horizontal lines: $\xi=0.24, 
0.49$, and $0.75$. The experimental data as well as the data for $\xi=1$, depicted also in 
the upper panel of Fig.~\ref{YieldFig}, are included for comparison. 
}
\label{YieldAngle}
\end{figure*}

Let us start with numerical data for
$W^\pm(E,\eta|E^\prime,\eta^\prime)=W^\pm_{\rm elastic}(E,\eta|E^\prime,\eta^\prime)
+W^\pm_{\rm impact}(E,\eta|E^\prime,\eta^\prime)$, the transition rates to be used
in~\eqref{GammaFct} for the calculation of the total scattering rate $\Gamma(E)$. The rates 
employed in~\eqref{Gfct} for the kernels $G^\pm(E,\eta|E^\prime,\eta^\prime)$ are the same
except of the factor two in front of $W^\pm_{\rm impact}(E,\eta|E^\prime,\eta^\prime)$
(see discussion in the paragraph after Eq.~\eqref{Wimpact} of the previous section). 

The angle dependence of $W^\pm(E,\eta|E^\prime,\eta^\prime)$ is depicted in 
Fig.~\ref{RatesAngleDependence} for the energy doublets $(E, E^\prime)=(28.8\,{\rm eV}, 28.8\,{\rm eV})$, 
representing elastic scattering, and $(28.8\,{\rm eV}, 1.8\,{\rm eV})$, standing for inelastic 
scattering. For the chosen energy, which is above the upper threshold  
$E^{\rm th}_2=3\,(2\pi/a)^2-\chi \simeq 10\,{\rm eV}$, elastic scattering arises solely 
from the ion cores. Due to the momentum dependence of the ion's pseudopotentials forward 
scattering is thus favored. Hence, $W^+(E,\eta|E,\eta^\prime)$, shown in the 
upper right panel, is peaked around $\eta^\prime=\eta$ and $W^-(E,\eta|E,\eta^\prime)$, 
plotted on the upper left, is large only for small $\eta$ and $\eta^\prime$, that is, for 
grazing scattering events. For initial and final state energies below the lower 
threshold $E^{\rm th}_1\simeq 1\,{\rm eV}$ (not shown), where in our model elastic scattering is due to phonons, 
both rates are isotropic because of the nonpolarity of the phonons, making the scattering 
strength $M^2$ in~\eqref{Wep} momentum independent. This energy range influences however the 
surface scattering kernel only indirectly via the total scattering rate $\Gamma(E)$ defined 
in~\eqref{GammaFct}, where the energy integral runs from $-\chi$ to $E$. Inelastic scattering in 
our model is due to impact ionization. From the panels of the second row of Fig.~\ref{RatesAngleDependence} 
it can be inferred that it does not favor any particular forward or backward direction. It is however 
again strongest in forward direction and there particularly for $\eta$ and $\eta^\prime$ 
close to unity.

Figure~\ref{RatesEnergyDependence} depicts the energy dependence of the transition 
rates for the direction cosine doublets $(\eta, \eta^\prime)=(1, 0.75)$ and $(1, 0.24)$. 
Elastic scattering, below $E^{\rm th}_1\simeq 1\,{\rm eV}$ due to phonons and above 
$E^{\rm th}_2\simeq 10\,{\rm eV}$ due to ion cores, is visible along the energy diagonal, 
whereas impact ionization leads to the off-diagonal data. Of particular relevance for the 
surface scattering kernel $R(E,\xi|E^\prime,\xi^\prime)$, and hence also for the secondary 
emission yield $Y(E,\xi)$, is the fact that impact ionization favors in backward direction 
for fixed initial and final direction cosines $\eta$ and $\eta^\prime$ large energy transfers. 
Hence, irrespective of the particular choice of $\eta$ and $\eta^\prime$, the rate 
$W^-(E,\eta|E^\prime,\eta^\prime)$ is strongest for small $E^\prime$. In forward 
direction small energy transfers also occur, leading to $W^+(E,\eta|E^\prime,\eta^\prime)$
covering larger parts of the $EE^\prime$ plane for a fixed pair $\eta$ and $\eta^\prime$.
However, the surface scattering kernel, as well as the emission yield, are determined by 
the interplay between forward and backward scattering. From the energy dependence of 
the transition rates $W^\pm(E,\eta|E^\prime,\eta^\prime)$ we thus expect that most of 
the backscattered and emitted electrons will have small energy. Hence, they will appear
close to the just-outside potential energy indicated by the dashed lines.

\begin{figure*}[t]
\begin{minipage}{0.5\linewidth}

\begin{minipage}{0.5\linewidth}
\rotatebox{270}{\includegraphics[width=0.85\linewidth]{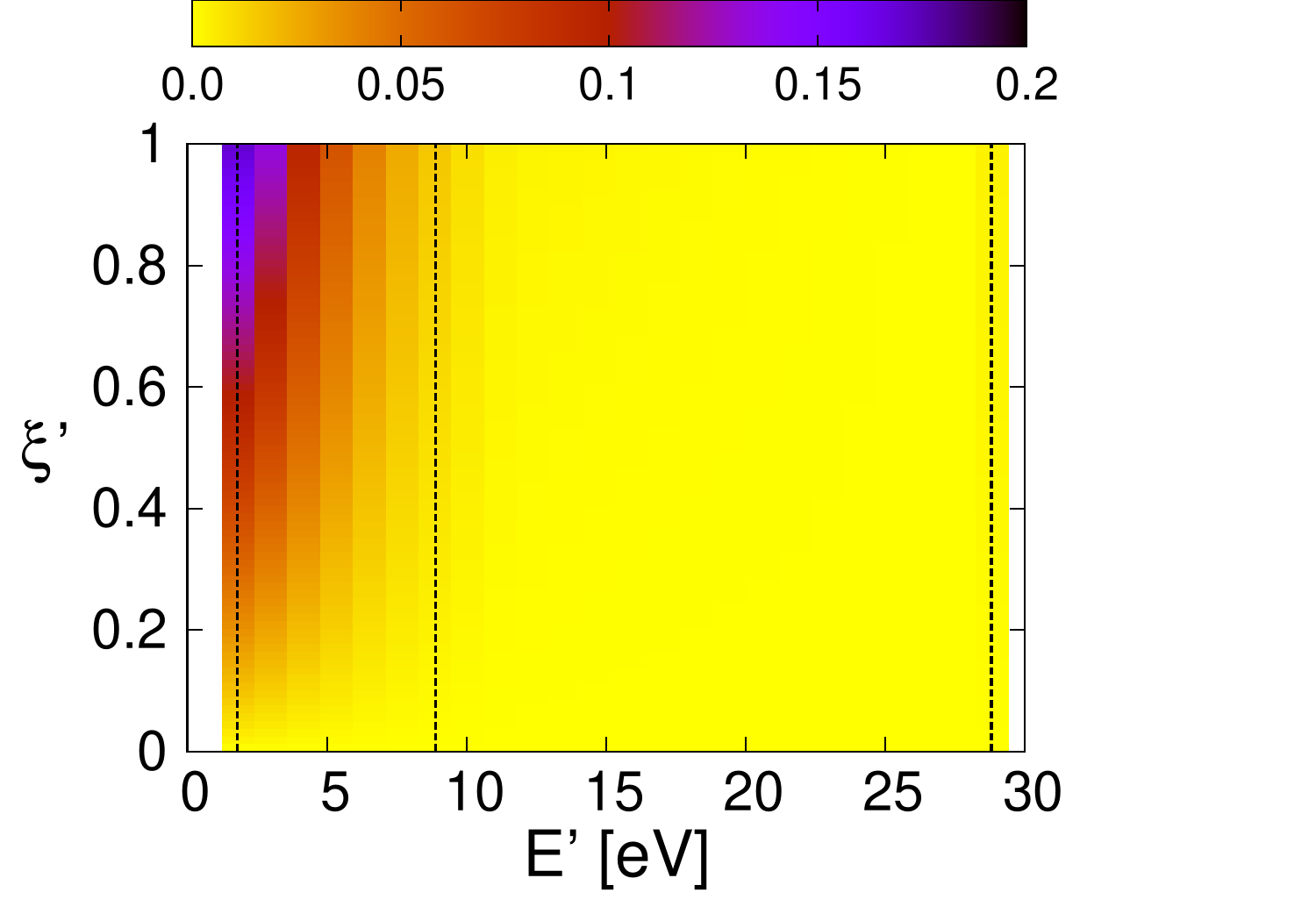}}

\rotatebox{270}{\includegraphics[width=0.85\linewidth]{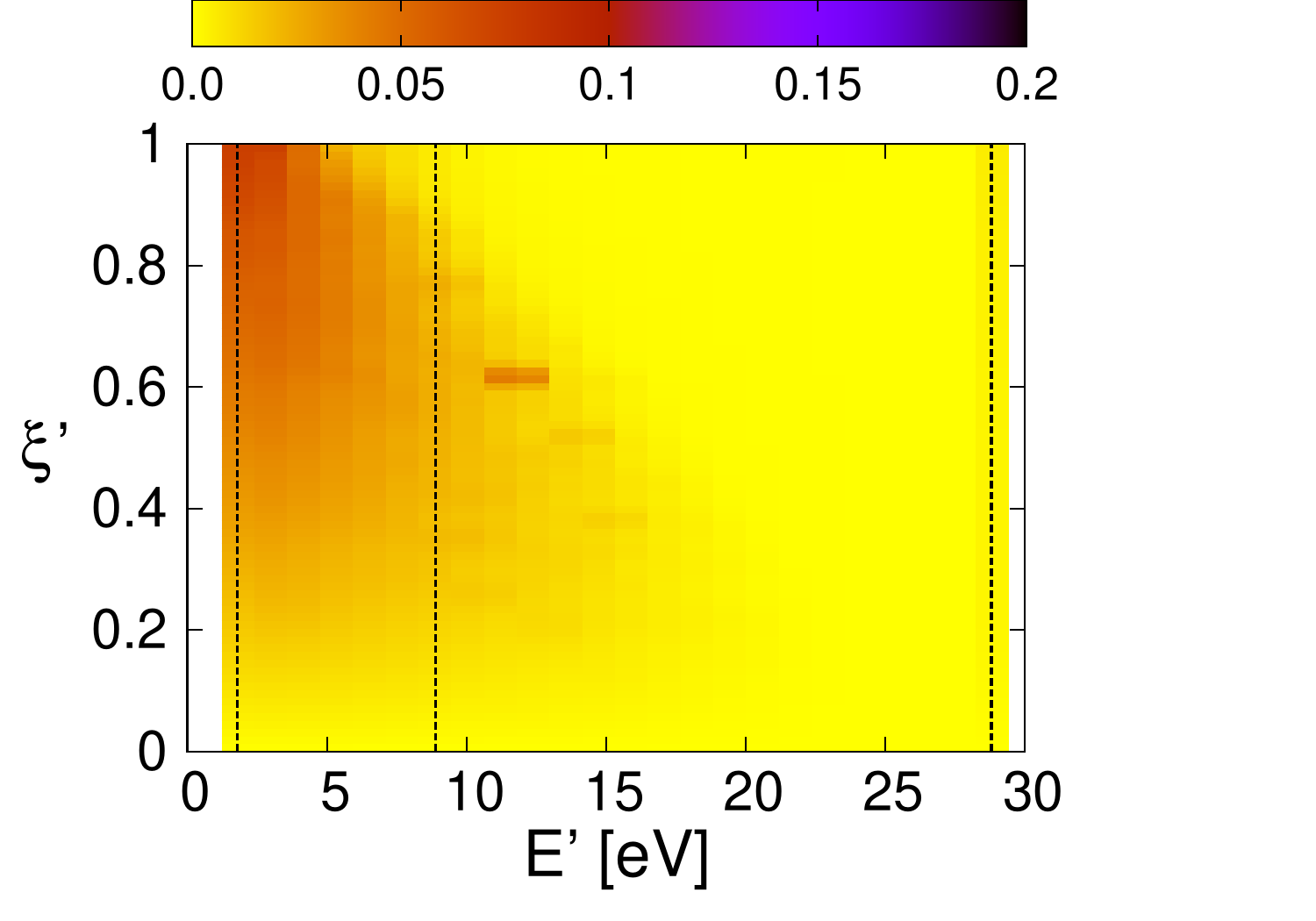}}
\end{minipage}\begin{minipage}{0.5\linewidth}
\rotatebox{270}{\includegraphics[width=0.85\linewidth]{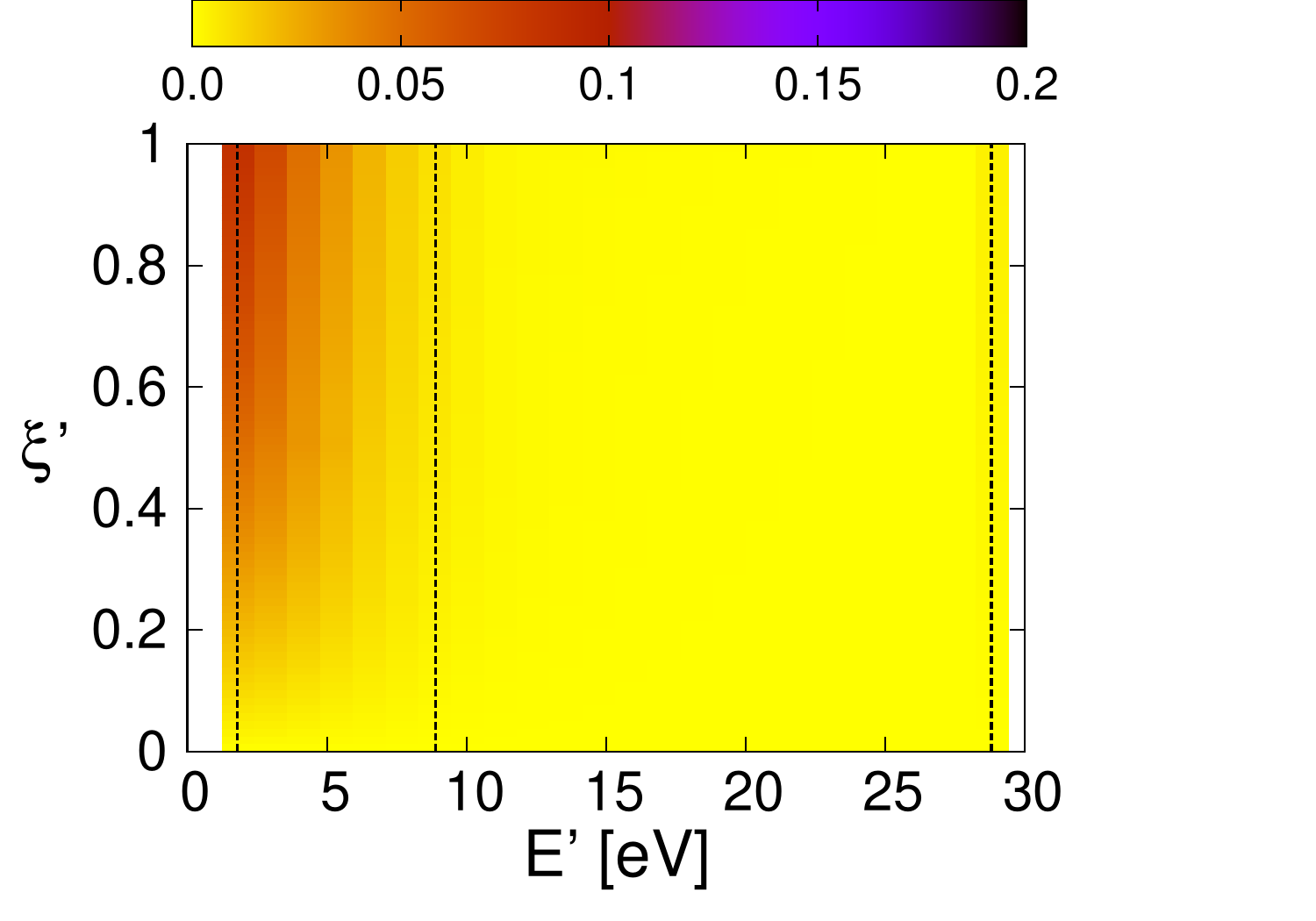}}

\rotatebox{270}{\includegraphics[width=0.85\linewidth]{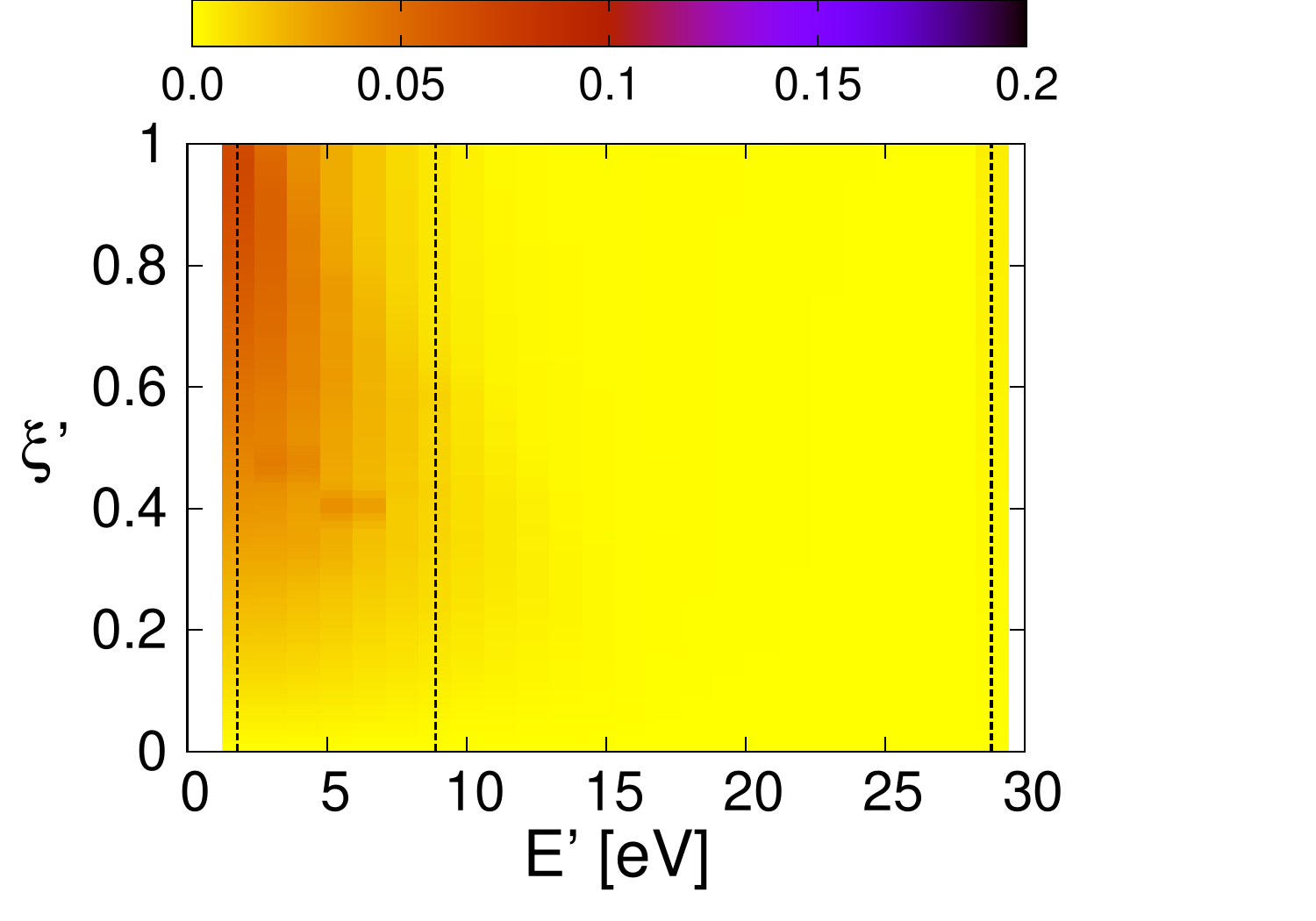}}

\end{minipage}
\end{minipage}\begin{minipage}{0.5\linewidth}
\hfill\includegraphics[width=0.965\linewidth]{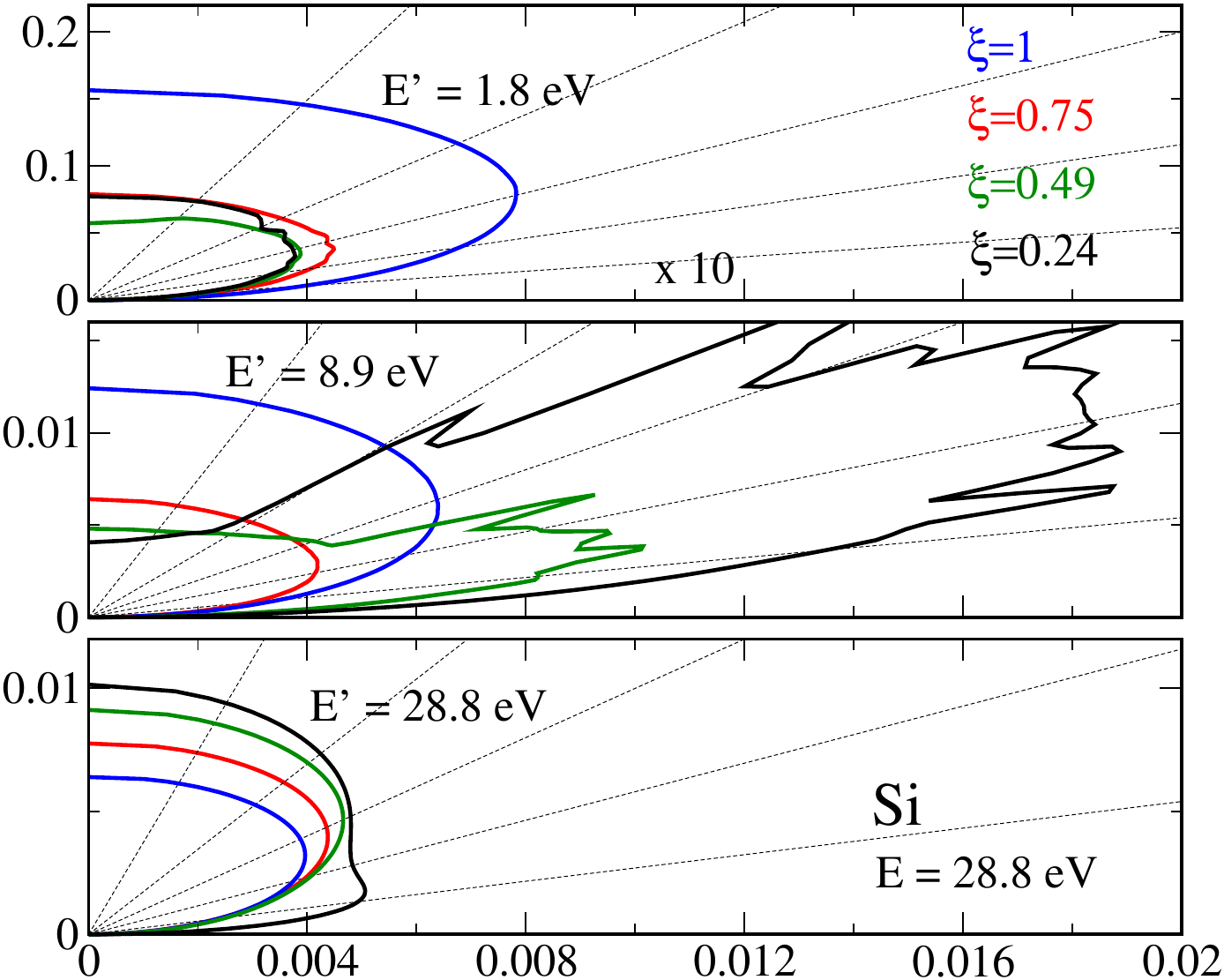}
\end{minipage}
\caption{(Color online) Left: Surface scattering kernel $R(E,\xi|E^\prime,\xi^\prime)$
for Si after a primary electron hit the surface with energy $E=28.8\,\mathrm{eV}$ and 
direction cosine $\xi=1, 0.75, 0.49,$ and $0.24$ (clockwise starting from the upper left). 
Right: Polar representation of the scattering kernel shown on the left for the three final
state energies indicated by the vertical dashed lines: $E^\prime=28.8\,\mathrm{eV}, 
8.9\,\mathrm{eV},$ and $1.8\,\mathrm{eV}$.
The rays indicate directions in steps of $\Delta\beta^\prime=\pi/12$. Notice the change of
scales in the two axes of the upper panel showing the data for $E^\prime=1.8\,{\rm eV}$. The
spikes in the data for $E^\prime=8.9\,{\rm eV}$ are of no physical relevance. They are due to
the noise of the Monte Carlo integration (required for the calculation of the impact ionization
transition rate) which is particularly strong for final state energies, where
$R(E,\xi|E^\prime,\xi^\prime)$ is rather small.
}
\label{SSKforSiData}
\end{figure*}

To validate at least partly the calculated transition rates 
$W^\pm(E,\eta|E^\prime,\eta^\prime)$ we show in Fig.~\ref{GammaFig} the total 
scattering rate $\Gamma(E)$ for silicon. Without the elastic scattering processes, 
$\Gamma(E)$ is the ionization rate and can be compared with the results of others. 
As discussed by Cartier and coworkers~\cite{CFE93}, there is a substantial spread 
in the calculated ionization rates. Below the just-outside potential energy, 
the rates are sensitive to details of the band structure. Since we 
are interested at energies above it, we do not enter this 
discussion. To demonstrate that the ionization rate of our model is plausible, 
we compare it specifically with the rates obtained by Bude and coworkers~\cite{BHI92} 
and Thoma and coworkers~\cite{TPE91}. Notice, both groups calculate the rate only
up to 3-4\,eV above the conduction band minimum, that is, below the 
just-outside potential energy. For impact ionization alone (ii), our data are sufficiently 
close to the data of the two groups, indicating that the semiconductor model we set up
produces reasonable data. Adding the elastic scattering on phonons and 
ion-cores (ii+ep/eic), with the linear switching between the two discussed above, 
modifies the rate. In particular the sharp on-set of impact ionization at 
$E \simeq -\chi+E_g \simeq -3\,{\rm eV}$ is 
whipped out and the rate assumes finite values all the way down to the bottom of the 
conduction band. More important for the surface scattering kernel is however the 
increase of $\Gamma(E)$ above the just-outside potential energy which can be also 
clearly seen. For comparison, we also plot in Fig.~\ref{GammaFig} the rates for 
electrons suffering in addition to impact ionization also elastic scattering on 
phonons (ii+ep) or ion cores (ii+eic) throughout the whole energy range. 

\subsection{Secondary electron emission yield}

The numerical data for $\Gamma(E)$ suggest that the model presented in~\ref{Model} is 
sufficiently close to reality to expect plausible emission yields and surface 
scattering kernels. Let us first look at the emission yields for which experimental
data exist. Figure~\ref{YieldFig} shows calculated emission yields for 
silicon and germanium together with experimental data from Bronshtein and 
Fraiman~\cite{BronFrai69} and Fowler and Farnsworth~\cite{FF58}. For silicon the 
agreement is satisfactory, given the fact that we do not adjust any parameter, working, 
for instance, with the pseudopotentials also employed in band structure 
calculations~\cite{ILC78,SCL75}. Our results for silicon are of the same quality as 
the Monte Carlo results of Pierron and coworkers~\cite{PIB17}, who used however a 
different model. For germanium the discrepancy between calculated and measured yields 
is larger. Only the order of magnitude is correct. Since silicon and germanium are similar 
materials, as can be seen from the material parameters, the failure for germanium is 
surprising. Further measurements as well as calculations are required to clarify the issue. 

We also plotted, again for comparison, the yields obtained by letting 
electron-phonon and electron-ion-core scattering, or both act throughout the whole energy 
range. Due to the isotropy of electron-phonon scattering in nonpolar semiconductors (see 
Eq.~\eqref{Wep}), the yield is substantially off the experimental data as soon as electron-phonon 
scattering is allowed throughout the whole energy range (ii+ep or ii+ep+eic). Allowing, on the 
other hand, electron-ion-core scattering throughout the whole energy range leads to a too small
emission yield. However, as discussed in~\ref{Model}, scattering on phonons should be dominant 
at low energy, whereas scattering on ion cores should be relevant at high energies. Indeed, 
switching from the former to the latter between $E^{\rm th}_1$ and $E^{\rm th}_2$, indicated 
in Fig.~\ref{YieldFig} by the dashed vertical lines, produces the best results (ii+ep/eic). 
This finding is also in accordance with Monte Carlo simulations of secondary electron emission 
which unisono incorporate incoherent scattering of electrons on ion cores, as we do, the only 
difference being in the choice of the scattering potential. Being mostly concerned with energies 
above $100\,{\rm eV}$ they employ screened atomic potentials (see, for instance, Pierron et 
al.~\cite{PIB17}, Werner~\cite{Werner23}, and Dapor's book~\cite{Dapor20}).

Of interest is also the angle dependence of the emission yield. Based on theoretical 
considerations~\cite{Dekker58} concerning the escape depth of electrons created by 
the primary electron, the emission yield for fixed energy is expected~\cite{Schou88,Tolias14a} 
to increase with decreasing direction cosine according to $Y(E,\xi)=Y(E,1)/\xi^n$, where 
$0< n <2$. For impact energies in the $\mathrm{keV}$ range, the emission yield indeed 
increases with decreasing $\xi$, supporting hence the cosine law and also the assumption
that most secondaries are created by the primary electron itself. In the 
very low energy range investigated in this work, however, the law is not valid. 
The data shown in Fig.~\ref{YieldAngle} do not support it. Instead, as can be most clearly
seen in the left panel, the emission yield has a weak nonmonotonous angle dependence, 
which moreover depends on the impact energy, suggesting that in this energy range electrons
escaping the solid are produced by a subtly interplay between the various scattering processes 
active inside the solid. Hence, depending on the individual strength of the processes, one or 
the other angle dependence may prevail and manifest itself in the angle dependence of the emission 
yield. 

\subsection{Electron surface scattering kernel}

Let us now take a look at representative data for the surface scattering kernel 
arising from our model for a silicon-plasma interface. It is this object 
which is most important for plasma modeling. Since for germanium the data 
are similar, we discuss only silicon. 

Figure~\ref{SSKforSiData} depicts on the left $R(E,\xi|E^\prime,\xi^\prime)$ for initial 
energy $E=28.8\,{\rm eV}$ and initial direction cosines $\xi=1, 0.75, 0.49,$ and $0.24$. 
As expected from the transition rates' favoring of large energy transfers, demonstrated in
Fig.~\ref{RatesEnergyDependence}, the surface scattering kernel is in all cases largest 
close to the just-outside potential energy. The polar plots (in $\beta^\prime$) of the kernel 
shown on the right of the figure indicate, moreover, that the emission occurs essentially
isotropically in all spatial directions compatible with the halfspace geometry, 
irrespective of the angle of incident. Only electrons emitted at intermediate energies, 
for instance, $E^\prime=8.9\,{\rm eV}$ show a preferred range of emission direction
in case of oblique incident. This can be also seen on the left, where for $\xi<0.75$ 
the kernel starts to reach out for oblique $\xi^\prime$ to larger final energies $E^\prime$. 
However, the magnitude of $R(E,\xi|E^\prime,\xi^\prime)$ is in this range of variables rather 
small. Hence, the directed emission is rather improbable. 
The main feature of the data shown on the left of Fig.~\ref{SSKforSiData}, that the kernel 
is almost vanishing for $8.9\,{\rm eV} < E^\prime < E$, remains intact by reducing $E$. 
Only for $E$ of a few eV the separation breaks down.

\begin{figure}[t]
\includegraphics[width=0.99\linewidth]{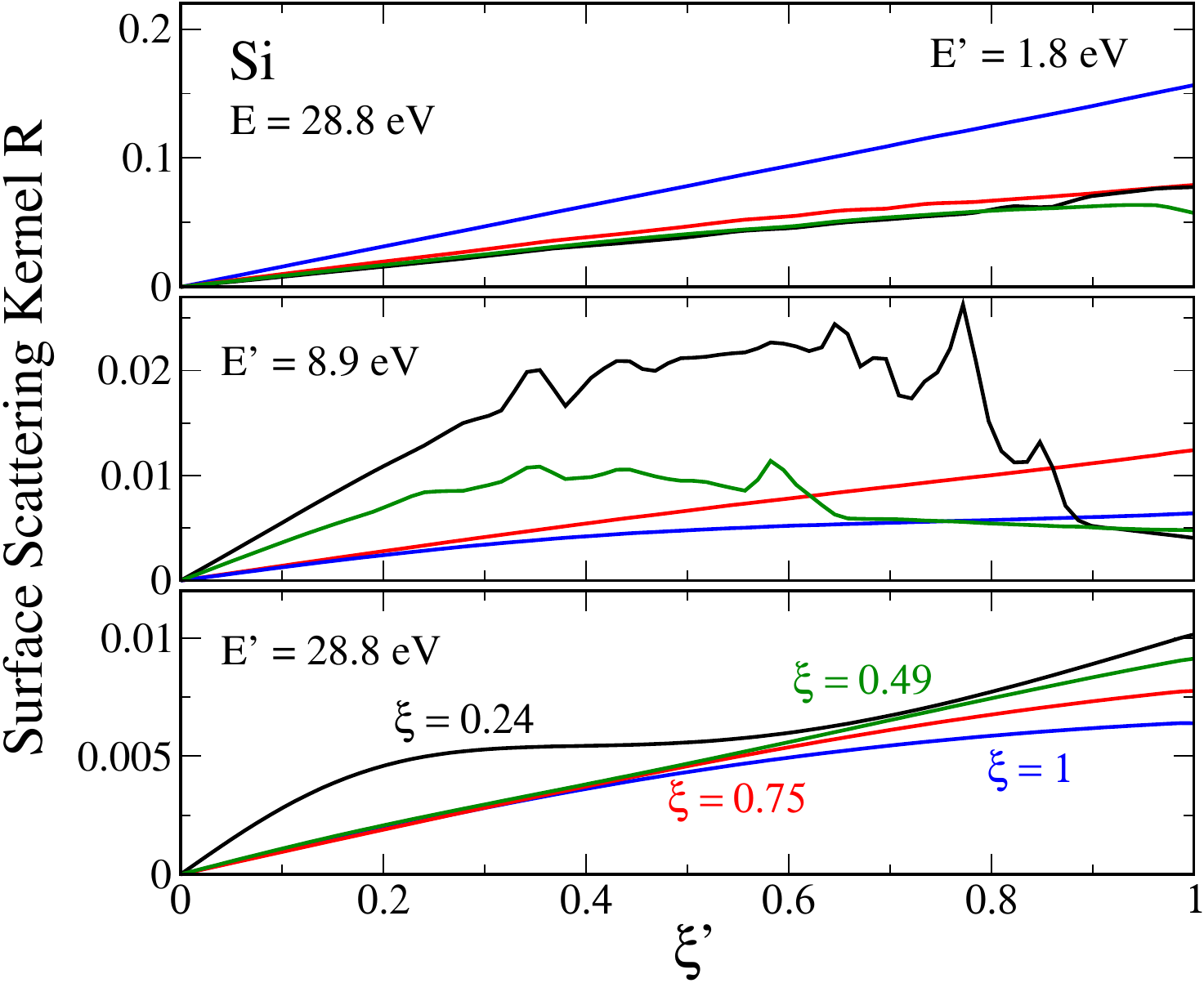}
\caption{(Color online) Variation of the surface scattering kernel $R(E,\xi|E^\prime,\xi^\prime)$
with the direction cosine $\xi$ for $E=28.8\,{\rm eV}$ and
$E^\prime=1.8, 8.9, 28.8\,{\rm eV}$. The peaks in the data for $E^\prime=8.9\,{\rm eV}$
and $\xi=0.49$ and $0.24$ are due to the noise of the Monte Carlo integration required for
computing the impact ionization transition rate. They have no physical relevance.
}
\label{SSKforSiDirection}
\end{figure}

Quantitatively, the dependencies on $E^\prime$ and $\xi^\prime$ can be better read off in 
Fig.~\ref{SSKforSiDirection}, where we plot some of the data for $R(E,\xi|E^\prime,\xi^\prime)$ 
in a different way. Notice, the scale of the ordinate of the upper panel, showing the data for 
$E^\prime=1.8\,{\rm eV}$, is a factor 10 larger than the corresponding one in the middle 
and lower panels, indicating that irrespective of the direction of impact $R(E,\xi|E^\prime,\xi^\prime)$ 
is in general largest close to the just-outside potential energy. Irrespective of $\xi$, the 
magnitude of $R(E,\xi|E^\prime,\xi^\prime)$ increases for most $\xi$ values monotonously with 
$\xi^\prime$. Only at intermediate emission energies $E^\prime\simeq 8.9\,{\rm eV}$ and oblique 
incident, for instance, for $\xi=0.49$ and $\xi=0.24$ develops $R(E,\xi|E^\prime,\xi^\prime)$ a very 
shallow maximum at finite $\xi^\prime$ and hence a preferred range of emission directions away from 
$\xi^\prime=1$, as already seen in Fig.~\ref{SSKforSiData}. The feature occurs, however, in an energy 
range, where $R(E,\xi|E^\prime,\xi^\prime)$ is rather small. Most secondary electrons are hence 
emitted perpendicularly close to the just-outside potential of the surface. Plots for other 
values of $E$, $\xi$, and $E^\prime$ show the same overall features.

To visualize the surface scattering kernel $R(E,\xi|E^\prime,\xi^\prime)$,
a function depending on four continuous variables, in its totality is of course 
impossible. Many more plots could be produced and put on display. They all 
look rather similar. For the purpose of demonstrating that the surface scattering 
kernel $R(E,\xi|E^\prime,\xi^\prime)$ can be obtained by the invariant embedding 
approach from a physical model of the plasma-facing solid the plots presented should 
suffice.

\section{Conclusion}
\label{Conclusions}

We presented a scheme for constructing a boundary condition for the electron energy
distribution function of a plasma in contact with a semiconducting solid. Based on 
the invariant embedding principle for the backscattering function, we derived an 
expression for the electron surface scattering kernel which takes the electron 
microphysics inside the solid, responsible for electron backscattering and emission, into 
account. The kernel connects at the plasma-solid interface the distribution function 
of the outgoing electrons with the distribution function of the incoming ones. Hence, 
an electron boundary condition arises which takes material-dependent aspects more 
faithfully into account as the phenomenological approaches used so far.

As an illustration, we applied the scheme to a silicon and a germanium surface, 
describing the electron's microphysics inside the solids by a randium-jellium 
model. Approximating the interface potential by a Schottky barrier and taking impact 
ionization as well as incoherent elastic scattering due to phonons and ion  
cores into account, we deduced from the surface scattering kernel emission yields 
in satisfactory agreement with experimental data. For silicon we obtained in 
fact good agreement, suggesting that the model captures the essential processes 
leading to electron backscattering and emission sufficiently well.

The approach quantifies the intrinsic electron-induced backscattering and emission 
properties of the wall material. Emission processes depending also on the plasma, 
for instance, field-assisted thermionic emission or emission due to impacting 
ions and radicals, have to be treated separately, if the need arises, and 
incorporated into the surface scattering kernel by additional terms.

A great advantage of the embedding approach is its numerical efficiency. We did not
benchmark the approach but it is out of question that building up the electron 
surface scattering kernel $R(E,\xi|E^\prime,\xi^\prime)$ by a Monte Carlo simulation,
for instance, would be rather costly since it requires statistical sampling for 
each doublet $(E,\xi)$ of initial energy and direction cosine, which, for the data we 
presented, implies around $2000$ individual Monte Carlo runs. Moreover, in contrast to 
approaches based on an electron Boltzmann equation for the solid halfspace, the embedding 
approach takes the electron microphysics responsible for intrinsic electron-induced 
backscattering and emission into account without tracing the electron distribution 
function across the plasma-solid interface. At least in situations where the emissive 
properties of the plasma-solid interface are not affected by the plasma, it is thus 
not necessary to run the computation of the surface scattering kernel together with 
the plasma simulation. The kernel can be computed before hand, stored in a data file, 
and read out in the boundary module of the plasma simulation without additional costs.

Whereas the transport problem in the form of the embedding equation is completely 
solved numerically, without linearization and also without an approximate decoupling
of direction cosines and energies, the randium-jellium modeling of the electron's microphysics 
inside the solid may not be the final answer, not only because of the incomplete agreement 
with experimental data for germanium, but also due to open conceptual issues. In 
particular, the scattering on the ion cores needs further studies. We assumed the 
scattering to be completely incoherent, making our model most appropriate for amorphous 
surfaces. In crystalline solids there should be however also coherent scattering,
leading to distinguished directions for backscattering and emission, as well as band gaps
above the just-outside potential energy. The mixed screening model, containing 
semiconductor and metal like elements, needs also further investigations. 

The randium-jellium model is however a good starting point for combining 
the electron kinetics of the solid with the one of the plasma. Provided 
pseudopotentials for the ion cores are available or can be constructed, it can be applied 
to other materials as well. Naturally, the scattering and screening processes have to be 
adjusted case by case, but in its main features, the model applies to semiconductors, 
dielectrics, as well as metals. Based on our previous work and the 
availability of material parameters, the randium-jellium model can be also applied to
Au, Ag, Cu, W, Al, \SiOTwo, and \AlTwoOThree\,. For these materials, the electron surface 
scattering kernel can hence be computed by the scheme layed out in the present
work.

The plausibility of the surface scattering kernels, and 
hence of the boundary conditions for the electron energy distribution functions, depends 
on how well the model captures the microscopic processes responsible for 
electron backscattering and emission. In order to determine the quality of the 
model, experimental input is critical. Not only measuring the emission yields 
and backscattering probabilities of freestanding surfaces is required, but 
also operando surface diagnostics of plasma-exposed surfaces, revealing the structural 
and chemical state of the surface hit by the electrons of the plasma, which 
could then be fed into the modeling of the plasma-solid interface.\\

We acknowledge support by the Deutsche Forschungsgemeinschaft through 
project 495729137.

\appendix 

\section{Numerical approach} 
\label{Numerics}

The numerical method we adopt for solving the embedding equation~\eqref{EmbeddingEq}
is due to Shimizu and coworkers~\cite{SM66a,SA72}, who used it for studying transport 
problems in nuclear reactor physics. It utilizes the Volterra-type structure of the 
energy integrals to transform~\eqref{EmbeddingEq} into a set of matrix equations
in the discretized direction cosines and turns out to be surprisingly efficient. 

As a first step, the energy space is split into windows of width $\Delta E$ and 
for each function $A(E,\eta|E^\prime,\eta^\prime)$ a set of functions
\begin{align}
A_{nm}(\eta|\eta^\prime)=\int_n dE \int_m dE^\prime A(E,\eta|E^\prime,\eta^\prime) f_n(E)~,
\end{align}
is introduced, where $\int_n dE =\int_{E_n}^{E_{n+1}}dE$ denotes the integration over
the $n^{\rm th}$ energy window and $f_n(E)=1/\Delta E$ is a weight function. In contrast to 
Shimizu and coworkers, the energy windows we employ have all the same width. 
Numbering the windows from low to high energy by $n=1,2,...,N$, and approximating 
$A(E,\eta|E^\prime,\eta^\prime)\approx A_{nm}(\eta|\eta^\prime)f_m(E^\prime)$ for 
$E$ and $E^\prime$ inside the windows labelled by $n$ and $m$, respectively, the embedding 
equation reduces to a set of integral equations in the direction cosines. In 
a straight manner, one obtains for $m=n$ 
\begin{align}
S_n \ast Q_{nn} + Q_{nn}\ast S_n &= G_{nn}^- + Q_{nn}\ast G_{nn}^- \ast Q_{nn} \nonumber\\
& + G_{nn}^+\ast Q_{nn} + Q_{nn}\ast G_{nn}^+ 
\label{Qnn}
\end{align} 
and for $m<n$ 
\begin{align}
S_n\ast Q_{nm} + Q_{nm}\ast S_m = K_{nm}^- 
\label{Qnm}
\end{align}
with 
\begin{align}
K_{nm}^- &= Q_{nn}\ast G_{nn}^- \ast Q_{nm} 
+ Q_{nm}\ast G_{mm}^- \ast Q_{mm}  \nonumber\\
&+ G_{nn}^+\ast Q_{nm} + Q_{nm}\ast G_{mm}^+ \nonumber\\
&+ G_{nm}^- + D_{nm}^- + C_{nm}^- + A_{nm}^+ + B_{nm}^+~, 
\label{Knm}
\end{align}
where we introduced a matrix notation in the direction cosines and the $\ast$ operation,
which is the $\circ$ operation~\eqref{CircDef} without the energy integration. In 
addition we defined the matrices
\begin{align}
A_{nm}^+ &= \sum_{l=m}^{n-1} G_{nl}^+\ast Q_{lm}~,\label{Aplus}\\
B_{nm}^+ &= \sum_{l=m+1}^n Q_{nl} \ast G_{lm}^+~,\\
C_{nm}^- &= \sum_{l=m+1}^{n-1}\sum_{p=m}^l Q_{nl} \ast G_{lp}^- \ast Q_{pm}~,\\
D_{nm}^- &= \sum_{p=m}^{n-1} Q_{nn} \ast G_{np}^- \ast Q_{pm}~.
\label{Dminus}
\end{align}

\begin{figure}[t]
\includegraphics[width=0.99\linewidth]{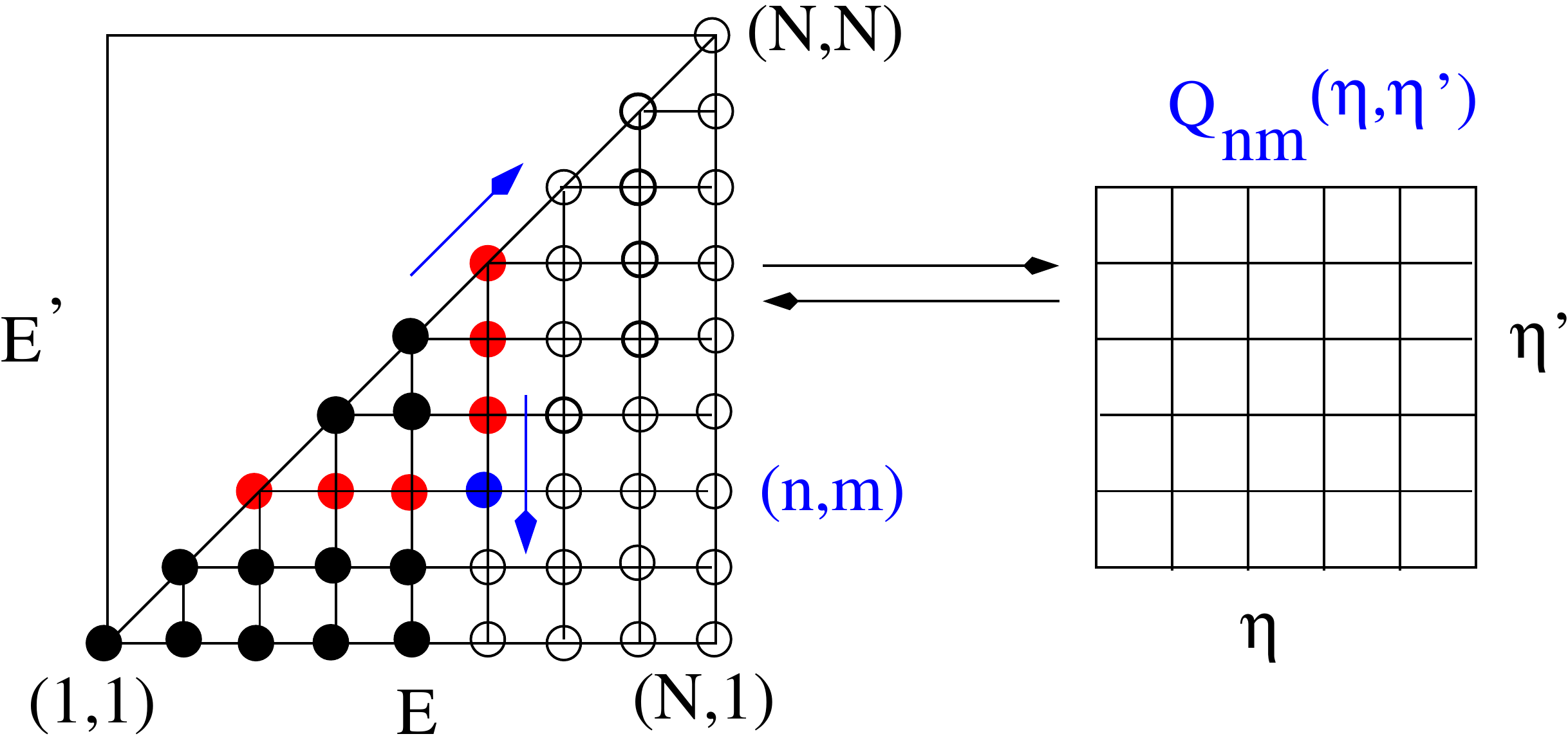}
\caption{(Color online) Numerical strategy for solving Eq.~\eqref{EmbeddingEq} for the function
$Q(E,\eta|E^\prime,\eta^\prime)$. Having discretized $Q(E,\eta|E^\prime,\eta^\prime)$
as described in the text, the algorithm calculates $Q_{nm}(\eta|\eta^\prime)$ on
an energy grid as indicated, starting with the diagonal $m=n$ and then working its
way through $m=n-r$ with $n=1,2,...,N$ and $r=1,2,...,n-1$. Black bullets indicate the
grid points for which $Q_{nm}(\eta|\eta^\prime)$ is known from the previous steps,
whereas open bullets indicate grid points not yet reached. The red bullets are
the grid points of the $Q_{lp}(\eta|\eta^\prime)$ entering the calculation of
$Q_{nm}(\eta|\eta^\prime)$ on the actual grid point $(n,m)$, shown as a blue bullet,
by iterating the discretized version of either~\eqref{Qnn}, if $m=n$, or~\eqref{Qnm},
if $m<n$, on the grid of discrete direction cosines depicted on the rhs of the figure.}
\label{NumStrategy}
\end{figure}

The crux is now to go through the energy space, that is, through the window indices 
in such a manner that all the matrices $Q_{kr}(\eta|\eta^\prime)$ appearing in
Eqs.~\eqref{Aplus}--\eqref{Dminus} are known from the previous steps of the calculation,
enabling thereby an iterative computation of the fixpoints $Q_{nn}$ and $Q_{nm}$ of, 
respectively, Eq.~\eqref{Qnn} and Eq.~\eqref{Qnm}.
As shown in Fig.~\ref{NumStrategy}, this is possible by first solving~\eqref{Qnn} for the 
diagonal elements $Q_{nn}$ and then solving~\eqref{Qnm} for the off-diagonal elements 
$Q_{nm}$ with $n>m$, where $m=n-r$ with $n=1,2,...,N$ and $r=1,2,...,n-1$. 

In a second step, the integrals over the direction cosine are discretized. Since we have 
to switch in the expression for the surface scattering 
kernel~\eqref{SSK} from internal ($\eta$) to external $(\xi$) 
direction cosines, we discretized the $\eta-$integrals by a Trapezian rule.
Interpolation enables us then to go from $\eta$ to $\xi$ and vice versa. 

At the end, the embedding equation~\eqref{EmbeddingEq} is thus turned into a set of matrix 
equations. To avoid matrix Ricatti and Sylvester equations for $Q_{nn}$ and $Q_{nm}$, 
respectively, it is advantageous to leave on the lhs of Eqs.~\eqref{Qnn} and~\eqref{Qnm} 
only the diagonal matrices $S_n$. It is then straightforward to iterate for the fixpoint 
matrices $Q_{nn}$ and $Q_{nm}$. For the results discussed in Sect.~\ref{Results} we split 
the energy range $[-\chi,E_{\rm max}]$ with $E_{\rm max}=30\,{\rm eV}$ into N=30 energy 
windows and used M=80 discretization points for the $\eta-$integrals. The function 
$Q(E,\eta|E^\prime,\eta^\prime)$, required for the surface scattering kernel, is finally 
approximated by
\begin{align}
\rho(E^\prime)Q(E,\eta|E^\prime,\eta^\prime)=\sqrt{\frac{\rho(E^\prime)}{\rho(E)}}Q_{nm}(\eta|\eta^\prime)f_m(E^\prime)~,
\end{align}
where $E$ and $E^\prime$ belong to the energy windows $n$ and $m$, respectively, 
and the factor involving the density of states arises from the 
symmetrization~\eqref{Symmetrization} which we adopted for a compact 
representation of the nonlinear term of the embedding equation. 

\section{Screening}
\label{Screening}

Avoiding a selfconsistent calculation of electric potentials inside the semiconductor, 
we adopt a phenomenological screening model which combines metal- and semiconductor-type 
screening. 

Following Phillips~\cite{Phillips68}, we split
the valence charge into an atomic part, localized close to
the ion cores, and a bond part, localized in the bonds between neighboring
ions. Anticipating a static dielectric constant $\varepsilon$,
denoting by $Z$ the valence of the atoms constituting the solid,
and measuring charge in units of the elementary charge $e$, the bond charge per
ion is $Z/\varepsilon$, while the atomic charge per ion is $Z(1-1/\varepsilon)$.
Assuming now, the atomic charges to perfectly screen the charge of the ions,
each ion core carries effectively only a charge $Z/\varepsilon$. Since the bond
charge is less localized then the atomic charge, it may approximately give rise
to metallic screening, described by a Thomas-Fermi screening wave number $k_s$
defined by
\begin{align}
k^2_s=\frac{12\pi n_b}{E_{F,b}}\,
\label{ks}
\end{align}
where $n_b=Zn_{\rm ion}/\varepsilon$ and $E_{F,b}=(3\pi^2n_b)^{2/3}$ is the Fermi
energy associated with the bond part of the valence charge density.

The screened Coulomb part of the ion's pseudopotential would thus become
$(Z/\varepsilon)/(q^2+k_s^2)$ with $k_s$ given by~\eqref{ks}. We have to correct
however $\varepsilon$ for the fact that part of the valence charge is put into
metallic screening. This can be done by using Penn's formula~\cite{Penn62} for
the dielectric constant of a semiconductor not for the total valence charge but
only for the atomic part. Hence, $Z$ in the Coulomb part of the pseudopotential
is not multiplied by $1/\varepsilon$ but by $1/\bar{\varepsilon}$ with
\begin{align}
\bar{\varepsilon}=1 + \frac{16\pi n_s}{(E^{\rm ave}_{\rm g})^2}\bigg(1-\frac{E_{\rm g}^{\rm ave}}{4 E_{F,s}}\bigg)~,
\label{PennFormula}
\end{align}
where $n_s=Z n_{\rm ion}(1-1/\varepsilon)$, $E_{F,s}=(3\pi^2n_s)^{2/3}$, and
$E_{\rm g}^{\rm ave}$ is the average optical band gap~\cite{Penn62,Phillips68}. Put
together, we then obtain the screened Coulomb part of the pseudopotential,
$(Z/\bar{\varepsilon})/(q^2+k_s^2)$, as it features in~\eqref{PseudoPotential}.

Since the impact ionization rate contains the Coulomb interaction between two
electrons, and not between an electron and an ion core, we screen it by the full
valence charge density in a metallic manner (see Appendix~\ref{ImpactRate}). Although this
is also an approximation, in spirit, it is consistent with previous calculations of the
impact ionization rate in semiconductors~\cite{Kane67,TPE91,BHI92,CFE93}.

\section{Impact ionization rate}
\label{ImpactRate}

Using spherical coordinates for the electron momenta $\vec{k}$ with the $z-$axis pointing 
into the solid and the standard Golden Rule expression~\cite{Kane67,TPE91,BHI92,CFE93}, 
the impact ionization rate due to scattering of a conduction band electron with 
momentum $\vec{k}$ to one with momentum $\vec{k}^\prime$ becomes, after one internal 
momentum integration is carried out, energies and lengths are measured in Rydbergs and 
Bohr radii, and electron and hole masses are set to the bare electron mass,
\begin{align}
{\cal W}(\vec{k}|\vec{k}^{\,\prime})=
\int d^3q\, |M(\vec{k},\vec{k}^{\,\prime},\vec{q})|^2
\Psi(E,-\tilde{E},E^\prime,E^{\rm CB}_{|\vec{q}^{\,\prime}|})~
\label{WimpactInitial}
\end{align}
with $\vec{q}^{\,\prime}=\vec{k}-\vec{k}^{\,\prime}+\vec{q}$. The squared matrix element 
for impact ionization reads in the approximation where overlap integrals between 
single-electron states are set to unity
\begin{align}
|M(\vec{k},\vec{k}^{\,\prime},\vec{q})|^2 &=2 \big( [U(\vec{k}-\vec{k}^\prime)]^2 +
[U(\vec{q}-\vec{k}^\prime)]^2 \nonumber\\
&- U(\vec{k}-\vec{k}^\prime) U(\vec{q}-\vec{k}^\prime) \big)
\end{align}
and the function taking care of the occupancy of the states in the conduction and 
valence band
\begin{align}
\Psi(E,-\tilde{E},E^\prime,E^{\rm CB}_{q^\prime})&=
\frac{2}{\pi^3}n_{\rm VB}(-\tilde{E})\bar{n}_{\rm CB}(E^{\rm CB}_{q^\prime})
\nonumber\\
&\times \delta(E-E^\prime-E_{q^\prime}^{\rm CB}-\tilde{E})~ 
\label{PsiFunction}
\end{align}
with $\bar{n}_{\rm CB}(E)=1-n_{\rm F}(E+\chi)$ and 
$n_{\rm VB}(-\tilde{E})=1-n_{\rm F}(\tilde{E}-E_g-\chi)$, where 
$n_{\rm F}(E)=1/(\exp(\beta E)+1)$ is the Fermi function.

In~\eqref{PsiFunction} we anticipated using the total energy $E$, the lateral kinetic 
energy $T=(E+\chi)(1-\eta^2)$ (or, equivalently, the direction cosine $\eta$), and the azimuth 
angle $\Phi$ as independent variables for the conduction band states and defined  
$E=E_k^{\rm CB}=k^2+T-\chi$ and $\tilde{E}=-E_{\vec{q}}^{\rm VB}=\vec{q}^{\,2}+\chi+E_g$, where the minus 
sign in front of $E_{\vec{q}}^{\rm VB}$ signals that $\tilde{E}$ denotes the energy of an hole in the 
valence band. The electron-electron interaction is given by 
\begin{align}
U(q)=\frac{1}{q^2+\kappa^2}
\end{align}
with $\kappa^2=12\pi n_{\rm t}/E_{F,t}$ the Thomas-Fermi screening wave number belonging to 
the total valence charge density $n_{\rm t}=Zn_{\rm ion}$ and $E_{F,t}=(3\pi^2n_t)^{2/3}$ is the 
Fermi energy associated with it, as discussed in Appendix~\ref{Screening}.

To proceed, the integration over $\vec{q}$ is transformed into an integration over $\tilde{E}$, 
$\tilde{T}$, and $\Phi_q$, where $\tilde{T}$ is the lateral kinetic energy of the valence 
band hole. Taking care of the Jacobi determinant associated with this variable transformation, 
measuring azimuth angles with respect to the projection of ${\vec k}$ onto the $xy-$plane,
which is the interface plane, and using the sign of the $z-$components of the momenta as labels 
to distinguish forwardly moving ($p=1$) from backwardly moving $(p=-1)$ electron states, the rate 
becomes after integrating out $\tilde{E}$ with the help of the energy-conserving 
$\delta$-function contained in~\eqref{PsiFunction},
\begin{widetext}
\begin{align}
{\cal W}(E,T,p|E^\prime,T^\prime,p^\prime)= 
\int_0^\infty \!\!\!d\tilde{T}\int_0^{2\pi}\!\!\!d\Phi_{k^\prime}\int_0^{2\pi}\!\!\! d\Phi_{q}
\sum_{i=1}^2 M_i(E,T,p|E^\prime,T^\prime,p^\prime;\tilde{T},\Phi_{k^\prime},\Phi_q)
\label{WimpactAppendix}
\end{align}
with 
\begin{align}
M_1(E,T,p|E^\prime,T^\prime,p^\prime;\tilde{T},\Phi_{k^\prime},\Phi_q) &= \sum_{\tilde{p}=\pm 1}
U(R_1,R_2)N(E,E^\prime,\tilde{E})\frac{\Theta(-c)}{\sqrt{r^2+8|c|}}\bigg|_{\tilde{q}=\tilde{q}^{(3)}_{\tilde{p}}}~,\\
M_2(E,T,p|E^\prime,T^\prime,p^\prime;\tilde{T},\Phi_{k^\prime},\Phi_q) &= \sum_{\tilde{p}=\pm 1}\sum_{j=1}^2
U(R_1,R_2)N(E,E^\prime,\tilde{E})
\frac{\Theta(c)\Theta(r^2-8c)}{\sqrt{r^2-8c}}\bigg|_{\tilde{q}=\tilde{q}^{(j)}_{\tilde{p}}}~,\\
U(R_1,R_2)&=2\bigg( [U(R_1)]^2 + [U(R_2)]^2 - U(R_1)U(R_2) \bigg)~,\\
N(E,E^\prime,\tilde{E})&= \pi^{-3} n_{\rm VB}(-\tilde{E})\bar{n}_{\rm CB}(E-E^\prime-\tilde{E})~,
\end{align}
and 
\begin{align}
R_1 &= |\vec{k}-\vec{k}^\prime|_{pp^\prime}=g(E,T,p|E^\prime,T^\prime,p^\prime;\Phi_{k^\prime})~,\\
R_2 &=  |\vec{q}-\vec{k}^\prime|_{\tilde{p}p^\prime}=
g(\tilde{E},\tilde{T},\tilde{p}|E^\prime,T^\prime,p^\prime;\Phi_q-\Phi_{k^\prime})~,\\
\tilde{E} &= E_g+\chi+\tilde{T}+\tilde{q}^2~,
\end{align}
where $\Theta(x)$ is the Heaviside step function and  
\begin{align}
\tilde{q}_{\tilde{p}}^{(1)}&=\frac{|r|}{4}\bigg(1+\sqrt{1-\frac{8c}{r^2}}\bigg)~,\\
\tilde{q}_{\tilde{p}}^{(2)}&=\frac{|r|}{4}\bigg(1-\sqrt{1-\frac{8c}{r^2}}\bigg)~,\\
\tilde{q}_{\tilde{p}}^{(3)}&=\frac{|r|}{4}\bigg(-\tilde{p}\,{\rm sign}(r)+\sqrt{1+\frac{8|c|}{r^2}}\bigg)~.
\end{align}
The functions $c$ and $r$ contained in $M_i(E,T,p|E^\prime,T^\prime,p^\prime;\tilde{T},\Phi_{k^\prime},\Phi_{q^\prime})$ 
are defined by 
\begin{align}
c(E,T,p|E^\prime,T^\prime,p^\prime;\tilde{T},\Phi_q,\Phi_{k^\prime}) &= 
\sqrt{\tilde{T}}\, h(T,T^\prime; \Phi_q,\Phi_q-\Phi_{k^\prime}) + 2\,\tilde{T} - E_g  
-s(E,T,p|E^\prime,T^\prime,p^\prime;\Phi_{k^\prime})~,\\
r(E,T,p|E^\prime,T^\prime,p^\prime) &= 2\, (p\sqrt{E+\chi-T} - p^\prime \sqrt{E^\prime + \chi -T^\prime})~,
\end{align}
where 
\begin{align}
h(T,T^\prime; \Phi_q,\Phi_q-\Phi_{k^\prime}) &= 2\, (\sqrt{T}\cos\Phi_q - \sqrt{T^\prime}\cos(\Phi_q-\Phi_{k^\prime}))~,\\
s(E,T,p|E^\prime,T^\prime,p^\prime;\Phi_{k^\prime}) &= E - E^\prime  - 
[g(E,T,p|E^\prime, T^\prime, p^\prime;\Phi_{k^\prime})]^2~,\\
g(E,T,p|E^\prime,T^\prime,p^\prime;\Phi_{k^\prime}) &= \big(T+T^\prime-2\,\sqrt{T T^\prime}\cos\Phi_{k^\prime}
+\big[p\sqrt{E+\chi-T}-p^\prime\sqrt{E^\prime + \chi - T^\prime}\big]^2\big)^{1/2}~.
\label{gtilde}
\end{align}
\end{widetext}
Due to the independent variables $E$, $T$, and $\Phi$ suggested by the interface geometry, the final 
expression for the impact ionization rate looks a bit messy. It follows however 
straight from energy and momentum conservation. The three remaining integrals  
in~\eqref{WimpactAppendix} have to be done numerically. For the data presented 
in Sect.~\ref{Results} we employed the Vegas Monte Carlo integrator of the 
Numerical Recipes~\cite{PTV96}.


\end{document}